\pdfoutput=1
\documentclass[twocolumn,showpacs,preprintnumbers,final,aps,citeautoscript,nofootinbib,pre,superscriptaddress,longbibliography,
]{revtex4-2}
\usepackage{amsmath}
\usepackage{amssymb}
\usepackage{amsfonts}
\usepackage{physics}
\usepackage{graphicx}
\graphicspath{{./figures/}}
\usepackage{float}
\usepackage{bm}
\usepackage{times}
\usepackage[colorlinks,bookmarks=true,citecolor=blue,linkcolor=blue,urlcolor=blue,breaklinks=true]{hyperref}
\usepackage{amsthm}
\usepackage{mathrsfs}
\usepackage{mleftright}
\usepackage{mathtools}
\usepackage{comment}
\usepackage{lmodern}
\usepackage{booktabs}
\usepackage{supertabular}
\usepackage{tikz-cd}
\usepackage{tikz}
\usetikzlibrary{matrix,decorations.pathreplacing}

\newcommand{\rmm}[1]{{\rm{#1}}}
\newcommand{\bmm}[1]{{\bm{#1}}}

\newcommand{\lr}[3] { \left#1 #2 \right#3}
\newcommand{\hc}[1]{ {#1} ^{\dagger} }

\newcommand{\com}[2]{\lr{[}{#1,#2}{]} }
\newcommand{\kitaiti}[1]{ \lr{\langle} {#1} {\rangle} }

\newcommand{\chuukakko}[1]{\lr{\{}{#1}{\}}}

\newcommand{\hf}{\frac{1}{2}}

\newcommand{\sx}{\sigma^{x}}
\newcommand{\sy}{\sigma^{y}}
\newcommand{\sz}{\sigma^{z}}

\newcommand{\im}{\rmm{i}}
\newcommand{\nt}{\notag \\}

\begin{document}
\title{Exact steady states and fragmentation-induced relaxation in the no-passing asymmetric simple exclusion process}
\preprint{Phys. Rev. E \textbf{113}, 064129 (2026); DOI: 10.1103/52vx-z447}
\author{Urei Miura} 
\email{urei.miura@yukawa.kyoto-u.ac.jp}
\affiliation{%
Division of Physics and Astronomy, Graduate School of Science, 
Kyoto University, Kyoto 606-8502, Japan}%
\affiliation{Center for Gravitational Physics and Quantum Information, Yukawa Institute for Theoretical Physics, 
Kyoto University, Kitashirakawa Oiwake-Cho, Kyoto 606-8502, Japan}%

\date{Received 29 April 2025; revised 3 April 2026; accepted 28 May 2026; published 15 June 2026}

\begin{abstract}
We introduce a multispecies generalization of the asymmetric simple exclusion process with a ``no-passing'' constraint, forbidding overtaking, on a one-dimensional reflecting (closed) chain without particle reservoirs.
This no-passing rule fragments the Hilbert space into an exponential number of disjoint sectors labeled by the particle species sequence, leading to relaxation dynamics that depend sensitively on the initial ordering.
We construct exact matrix-product steady states in every particle species sequence sector and derive closed-form expressions for the particle-number distribution and two-point particle correlation functions.
In the two-species case, we identify a parameter regime where some sectors have finite relaxation times while others exhibit metastable relaxation dynamics, revealing the coexistence of fast and slow dynamics and strong dependence on the particle species sequence sector.
Our results uncover a novel mechanism for nonequilibrium metastability arising from Hilbert space fragmentation in exclusion processes.
\end{abstract}

\maketitle

\section{Introduction}\label{sec:introduction}
Understanding many-body systems far from equilibrium is one of the central challenges in modern physics.
Because this problem is generally difficult, exactly solvable nonequilibrium many-body models are especially valuable.
A common strategy is therefore to analyze simple interacting lattice gases that still capture essential nonequilibrium transport and relaxation.
Within this class, the asymmetric simple exclusion process (ASEP) has become a paradigmatic benchmark and has long served as a cornerstone of nonequilibrium statistical physics~\cite{macdonald1968kinetics,SPITZER1970246,liggett1985interacting}.
In its minimal form, ASEP describes stochastic hopping under hard-core exclusion, with applications ranging from traffic flow~\cite{CHOWDHURY2000199} to biological transport~\cite{Chou_2011}.
A key feature is integrability rooted in quantum-group symmetry, which enables analysis via the Bethe ansatz and exact steady states via matrix-product methods~\cite{DERRIDA199865,Blythe_2007}.
Duality relations (see Ref.~\cite{borodin2014duality}) and connections to Kardar-Parisi-Zhang universality~\cite{HALPINHEALY1995215} have also attracted sustained interest.

In this work, we extend the ASEP to include $M$ species of particles and impose a ``no-passing'' constraint, meaning particles cannot overtake each other, drastically altering the model's relaxation dynamics.
The present model corresponds to Solution CII in the ASEP classification of Ref.~\cite{Blythe_2007}.
Its behavior under periodic boundary conditions was examined in Ref.~\cite{M.R.Evans_1996}, and the open-boundary totally asymmetric simple exclusion process (TASEP) limit was analyzed in Ref.~\cite{rakos2005bethe}.
Here we study ASEP with reflecting (closed) boundary conditions to clarify congestion patterns that emerge once the complete boundary jam of the TASEP limit is relaxed.
This ``no-passing'' model can be mapped onto a frustration-free quantum many-body system, enabling all steady states to be obtained exactly and explicitly.
Similarly to the Hubbard model with infinite on-site repulsion~\cite{PhysRevB.41.2326,PhysRevLett.134.010411} and the $t$-$J_{z}$ model~\cite{PhysRevB.101.125126}, the no-passing constraint leads to an exponential fragmentation of the Hilbert space into exponentially many sectors, a phenomenon known as Hilbert space fragmentation (HSF)~\cite{PhysRevX.10.011047,Moudgalya_2022}.
As a result, the relaxation dynamics do not converge to a single ``typical'' state but remain strongly dependent on the initial condition.

The structure of this paper is as follows.
In Sec.~\ref{sec:model}, we introduce the multispecies no-passing ASEP, formulate its master equation, and perform the imaginary-gauge similarity transformation to obtain a Hermitian Hamiltonian.
In Sec.~\ref{sec:symmetry} we detail the various symmetries of the model, including global $\text{U}\qty(1)$ conservation, parity, and species-swap unitaries, and the particle species sequence-preserving operator responsible for Hilbert space fragmentation.
In Sec.~\ref{sec:exact-steady-complexity}, we construct exact steady states from the detailed-balance condition and evaluate the computational complexity of direct expectation-value calculations in configuration space.
In Sec.~\ref{sec:ground_state}, we show that each particle species sequence sector has a unique ground state from the Markov process structure, and we exhibit the frustration-free nature.
Section~\ref{sec:statistical-mps} constructs exact matrix-product steady states of the non-Hermitian Markov generator and derives the normalization in closed form.
In Sec.~\ref{subsec:distribution-correlation} we develop the transfer-matrix method to compute exact one- and two-point density correlations.
In Sec.~\ref{sec:gap}, we analyze the many-body spectral gap in the two-species system and present exact results together with numerical evidence for three regimes: a finite-gap regime that controls fast relaxation, a gapless regime where relaxation slows down algebraically with system size and a regime in which the gap is exponentially small in system size, giving rise to metastable dynamics.
Finally, Sec.~\ref{sec:conclusion} summarizes our main findings and outlines several avenues for future work.
Appendix~\ref{app:symbols} provides a quick-reference table of symbols and definitions used throughout the paper.
Appendix~\ref{app:sector} shows how single-species sectors reduce to the standard ASEP and map to the Ising-like ferromagnetic XXZ chain.
Appendix~\ref{app:algebraic} presents an algebraic steady state construction via quantum-group-like raising operators.
Appendix~\ref{app:gappedproof} provides Knabe-threshold checks and numerical evidence for a finite bulk gap at sampled parameter points.
Appendix~\ref{app:expsmallnumerical} collects detailed scaling data for the minimal-gap sector.
Appendix~\ref{app:Hamrep} gives the explicit matrix representation used in our exact diagonalization studies.

\section{Model: no-passing ASEP}\label{sec:model}

\begin{figure}[H]
  \includegraphics[width=\columnwidth,clip]{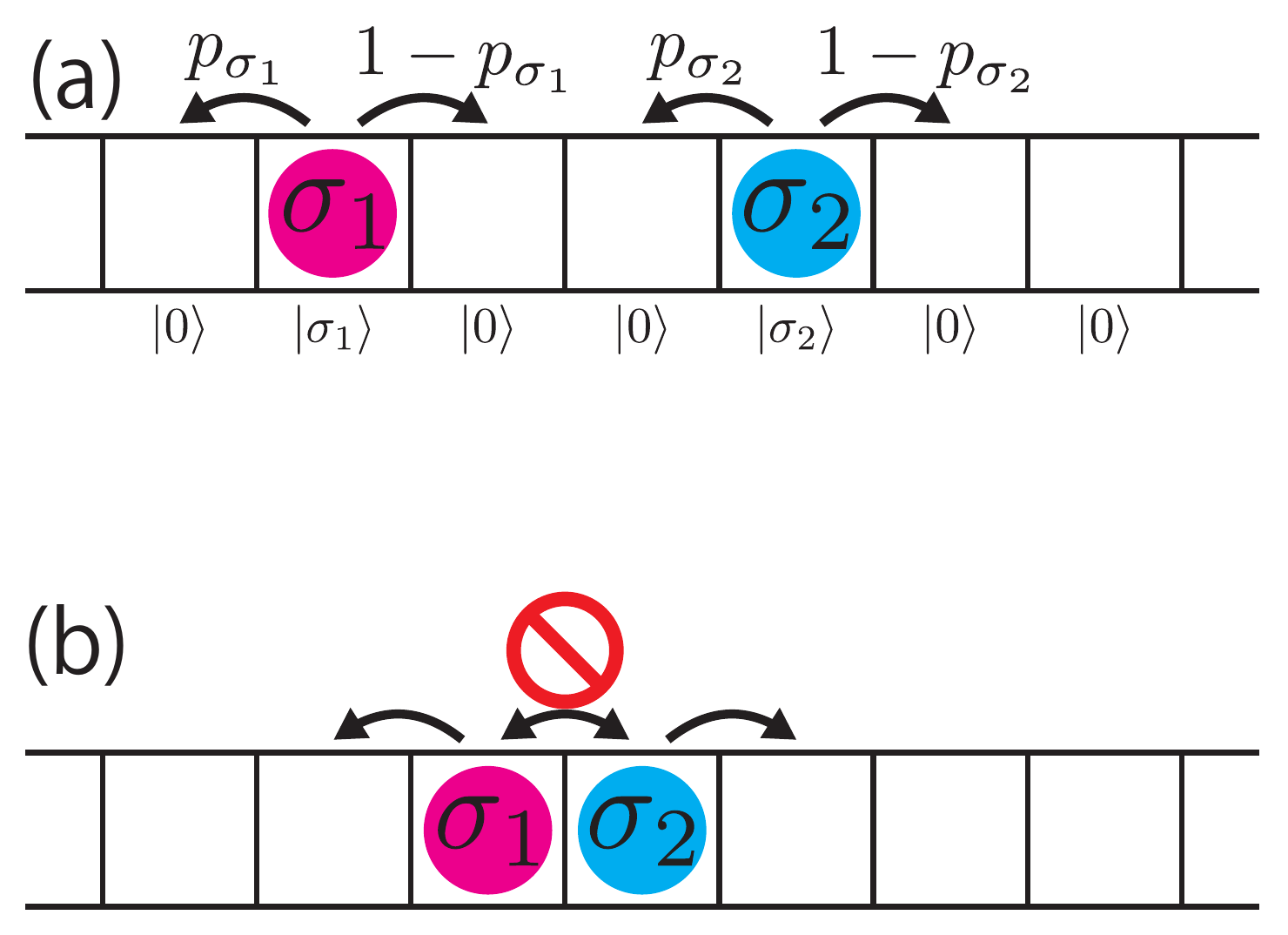}
  \caption{Schematic of the no-passing ASEP rules in reflecting (closed) boundary conditions.
  (a) A particle hops only when the adjacent site is empty.
  (b) Hopping into an occupied site is forbidden, so overtaking is not allowed.
  \label{fig:01}}
\end{figure}

In this section, we set up the event-space and Hilbert-space notation and then formulate the master equation of the multispecies no-passing ASEP.
We consider a one-dimensional lattice of length $L$ with reflecting (closed) boundary conditions (RBC), i.e., without particle reservoirs.
Each site obeys hard-core exclusion and is either empty or occupied by one particle among $M$ species.
Throughout this paper, on the bond $(j,j+1)$, $p_{\sigma}$ denotes the left-hop rate $\ket{0,\sigma}_{j,j+1}\to\ket{\sigma,0}_{j,j+1}$ and $1-p_{\sigma}$ denotes the right-hop rate $\ket{\sigma,0}_{j,j+1}\to\ket{0,\sigma}_{j,j+1}$; overtaking is forbidden (Fig.~\ref{fig:01}).

This model is a special case of Solution CII in~\cite{Blythe_2007}.
The periodic boundary condition (PBC) case was studied in~\cite{M.R.Evans_1996}, while the open-boundary TASEP case was studied in~\cite{rakos2005bethe}.
In this paper, we consider ASEP under reflecting (closed) boundary conditions.
This setup highlights congestion patterns that are rare under periodic boundaries and that collapse to a simple boundary pileup in the TASEP limit.

\subsection{Event and Hilbert space setup}\label{subsec:symbols}

We consider a one-dimensional lattice with $L$ sites and $M$ particle species under the hard-core constraint (at most one particle per site).
For a fixed particle number $N$ ($0\le N\le L$), a configuration is fully specified by ordered particle positions and species labels:
\begin{align}
  &(j_{1},\ldots,j_{N})\in \{1,\ldots,L\}^{\times N},
  \quad (j_{1}<\cdots<j_{N}),\\
  &(\sigma_{1},\ldots,\sigma_{N})\in \{1,\ldots,M\}^{\times N}.
\end{align}
Symbolically, we write
\begin{align}
  C\coloneqq ((j_{1},\sigma_{1}),\ldots,(j_{N},\sigma_{N})),
  \quad (j_{1}<\cdots<j_{N}).
\end{align}
The full set of $C$ defines the sample space, whose cardinality is
\begin{align}
  \sum_{N=0}^{L}\binom{L}{N}M^{N}=(M+1)^{L}.
\end{align}

To formulate the master equation in vector form, we define at each site $j\in\{1,\ldots,L\}$ an $(M+1)$-dimensional local Hilbert space $\mathcal{H}_{j}$ with basis states $\ket{0}_{j}$ (empty) and $\ket{\sigma}_{j}$ ($\sigma\in\{1,\ldots,M\}$):
\begin{align}
  \mathcal{H}_{j}=\mathrm{span}\qty(\ket{0}_{j},\ket{1}_{j},\ldots,\ket{M}_{j}).
\end{align}
As ladder operators between these states, we define hard-core boson operators
\begin{align}
  b_{j,\sigma}\coloneqq \ket{0}\bra{\sigma}_{j},\quad
  \hc{b}_{j,\sigma}\coloneqq \ket{\sigma}\bra{0}_{j}.
\end{align}
They satisfy the hard-core constraints
\begin{align}
  b_{j,\sigma}b_{j,\sigma'}=0,\quad
  \hc{b}_{j,\sigma}\hc{b}_{j,\sigma'}=0,
\end{align}
and act as
\begin{equation}
\begin{aligned}
  &b_{j,\sigma}\ket{0}_{j}=0,\quad
  b_{j,\sigma}\ket{\sigma'}_{j}=\delta_{\sigma,\sigma'}\ket{0}_{j},\\
  &\hc{b}_{j,\sigma}\ket{0}_{j}=\ket{\sigma}_{j},\quad
  \hc{b}_{j,\sigma}\ket{\sigma'}_{j}=0.
\end{aligned}
\end{equation}
For convenience, we define
\begin{align}
  \hat{n}_{j,\sigma}\coloneqq\ket{\sigma}\bra{\sigma}_{j}=\hc{b}_{j,\sigma}b_{j,\sigma},\quad
  \hat{n}_{j}\coloneqq\sum_{\sigma=1}^{M}\hat{n}_{j,\sigma}=1-\ket{0}\bra{0}_{j}.
\end{align}
These are diagonal operators with
\begin{equation}
\begin{aligned}
  &\hat{n}_{j,\sigma}\ket{0}_{j}=0,\quad
  \hat{n}_{j,\sigma}\ket{\sigma'}_{j}=\delta_{\sigma,\sigma'}\ket{\sigma'}_{j},\\
  &\hat{n}_{j}\ket{0}_{j}=0,\quad
  \hat{n}_{j}\ket{\sigma'}_{j}=\ket{\sigma'}_{j}.
\end{aligned}
\end{equation}
The full Hilbert space is
\begin{align}
  \mathcal{H}\coloneqq\bigotimes_{j=1}^{L}\mathcal{H}_{j},
\end{align}
and the empty state is
\begin{align}
  \ket{\Omega}\coloneqq\bigotimes_{j=1}^{L}\ket{0}_{j}=\ket{0,\ldots,0}.
\end{align}
This setup gives a one-to-one correspondence between sample-space elements and the orthonormal basis of $\mathcal{H}$: for $C\coloneqq((j_{1},\sigma_{1}),\ldots,(j_{N},\sigma_{N}))$ with $(j_{1}<\cdots<j_{N})$,
\begin{align}
  \ket{C}\coloneqq \hc{b}_{j_{1},\sigma_{1}}\cdots\hc{b}_{j_{N},\sigma_{N}}\ket{\Omega}.
\end{align}
Because hard-core boson operators on different sites commute, this definition is well posed.
Below, we identify $C$ and $\ket{C}$ when no confusion arises.

\subsection{Master equation and construction of the no-passing ASEP}\label{subsec:master}

We now fix notation for a finite-state continuous-time Markov process.
Let $\pi(t;C)$ be the probability of configuration $C$ at time $t$.
For transitions $C'\leftarrow C$ with $C\neq C'$, we write the rate as $-W_{C'\leftarrow C}$ ($W_{C'\leftarrow C}\le0$).
The diagonal total transition rate is defined by
\begin{align}
  -W_{C\leftarrow C}\coloneqq \sum_{C'(\neq C)} W_{C'\leftarrow C},
\end{align}
so $W_{C\leftarrow C}\ge0$ and probability conservation gives
\begin{align}
  \sum_{C'} W_{C'\leftarrow C}=0.
\end{align}
Formally, we define
\begin{align}
  &\ket{P(t)}\coloneqq \sum_{C}\pi(t;C)\ket{C},\\
  &\braket{C}{P(t)}=\pi(t;C),
\end{align}
and the transition matrix
\begin{align}
  &W\coloneqq\sum_{C,C'}W_{C'\leftarrow C}\ket{C'}\bra{C},\\
  &\bra{C'}W\ket{C}=W_{C'\leftarrow C}.
\end{align}
Then the master equation reads
\begin{align}
  \frac{\rmm{d}}{\rmm{d}t}\pi(t;C)=-\sum_{C'}W_{C\leftarrow C'}\pi(t;C'),
\end{align}
or, in vector form,
\begin{align}
  \frac{\rmm{d}}{\rmm{d}t}\ket{P(t)}=-W\ket{P(t)}.
\end{align}

We construct the no-passing ASEP by specifying local transitions on each bond $(j,j+1)$ ($j=1,\ldots,L-1$).
A particle of species $\sigma\in\{1,\ldots,M\}$ can hop only if the neighboring site is empty:
\begin{align}
  &\ket{0,\sigma}_{j,j+1}\to\ket{\sigma,0}_{j,j+1}:p_{\sigma},\\
  &\ket{\sigma,0}_{j,j+1}\to\ket{0,\sigma}_{j,j+1}:1-p_{\sigma},\\
  &\ket{\sigma',\sigma}_{j,j+1}\to\ket{\sigma,\sigma'}_{j,j+1}:0.
\end{align}
Thus, particle exchange (overtaking) is forbidden.
Using hard-core boson operators,
\begin{equation}
\begin{aligned}
  &\ket{0,\sigma}\bra{\sigma,0}_{j,j+1}=b_{j,\sigma}\hc{b}_{j+1,\sigma},\\
  &\ket{\sigma,0}\bra{0,\sigma}_{j,j+1}=\hc{b}_{j,\sigma}b_{j+1,\sigma},\\
  &\ket{0,\sigma}\bra{0,\sigma}_{j,j+1}=(1-\hat{n}_{j})\hat{n}_{j+1,\sigma},\\
  &\ket{\sigma,0}\bra{\sigma,0}_{j,j+1}=\hat{n}_{j,\sigma}(1-\hat{n}_{j+1}),
\end{aligned}
\end{equation}
and therefore the local transition matrix is
\begin{align}
\label{eq:nonHHam}
  W_{j,j+1}
  =&-\sum_{\sigma=1}^{M}\qty[p_{\sigma}\hc{b}_{j,\sigma}b_{j+1,\sigma}
  +(1-p_{\sigma})\hc{b}_{j+1,\sigma}b_{j,\sigma}]\nt
  &+\sum_{\sigma=1}^{M}\qty[p_{\sigma}(1-\hat{n}_{j})\hat{n}_{j+1,\sigma}
  +(1-p_{\sigma})\hat{n}_{j,\sigma}(1-\hat{n}_{j+1})].
\end{align}
Hence the full transition matrix is
\begin{align}
  W=\sum_{j=1}^{L-1}W_{j,j+1}.
\end{align}

\begin{figure}[H]
  \includegraphics[width=\columnwidth,clip]{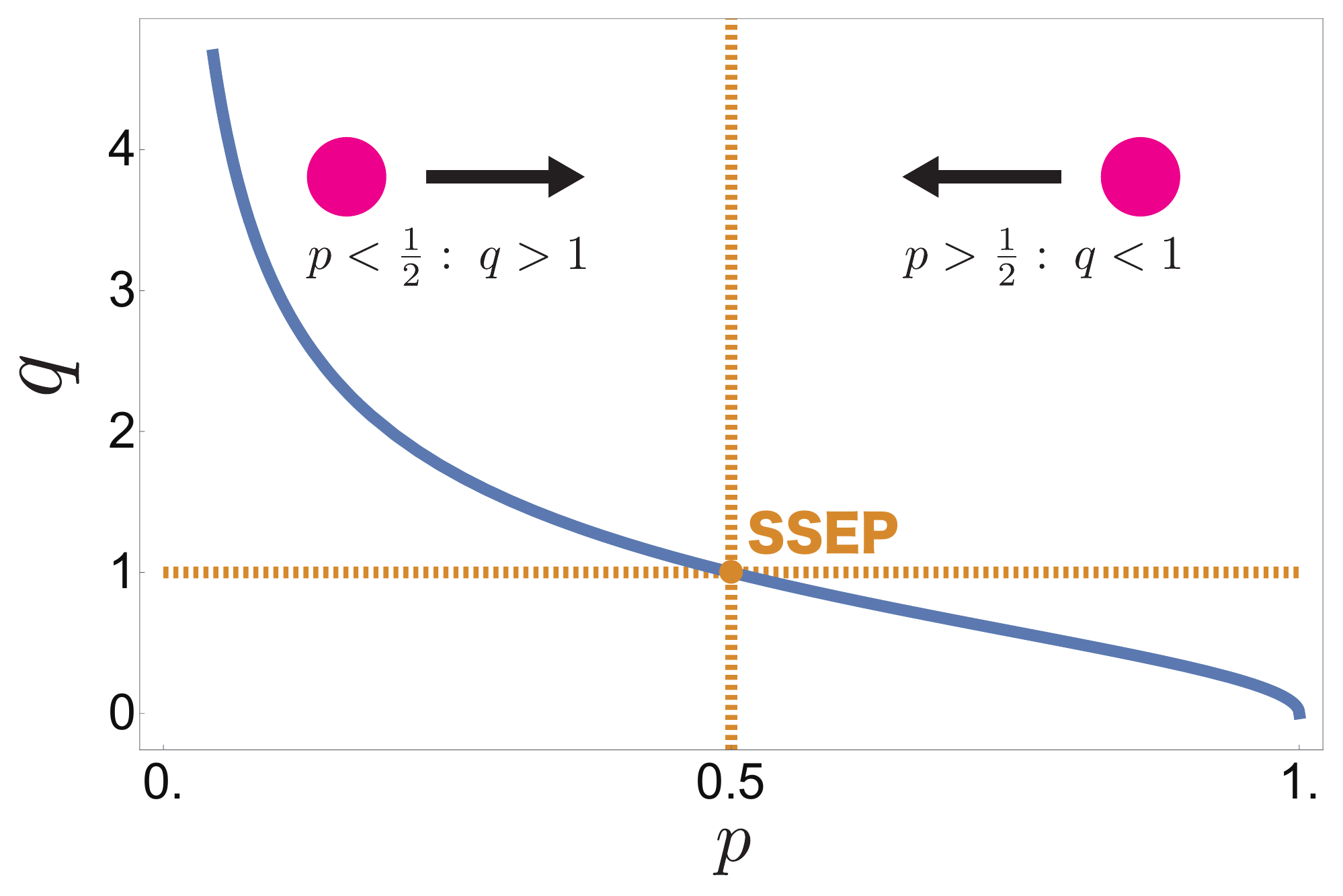}
  \caption{Relation between $p$ and $q$.
  For $0<p<1/2$, the bias is to the right ($q>1$). For $1/2<p<1$, the bias is to the left ($q<1$). At $p=1/2$ (symmetric simple exclusion process, SSEP case), $q=1$. 
  \label{Fig-p-qplot}}
\end{figure}

For later use, we define
\begin{align}
  q_{\sigma}\coloneqq\sqrt{\frac{1-p_{\sigma}}{p_{\sigma}}}.
\end{align}
Figure~\ref{Fig-p-qplot} shows the relation between $p_{\sigma}$ and $q_{\sigma}$.

\begin{figure}[H]
  \includegraphics[width=\columnwidth,clip]{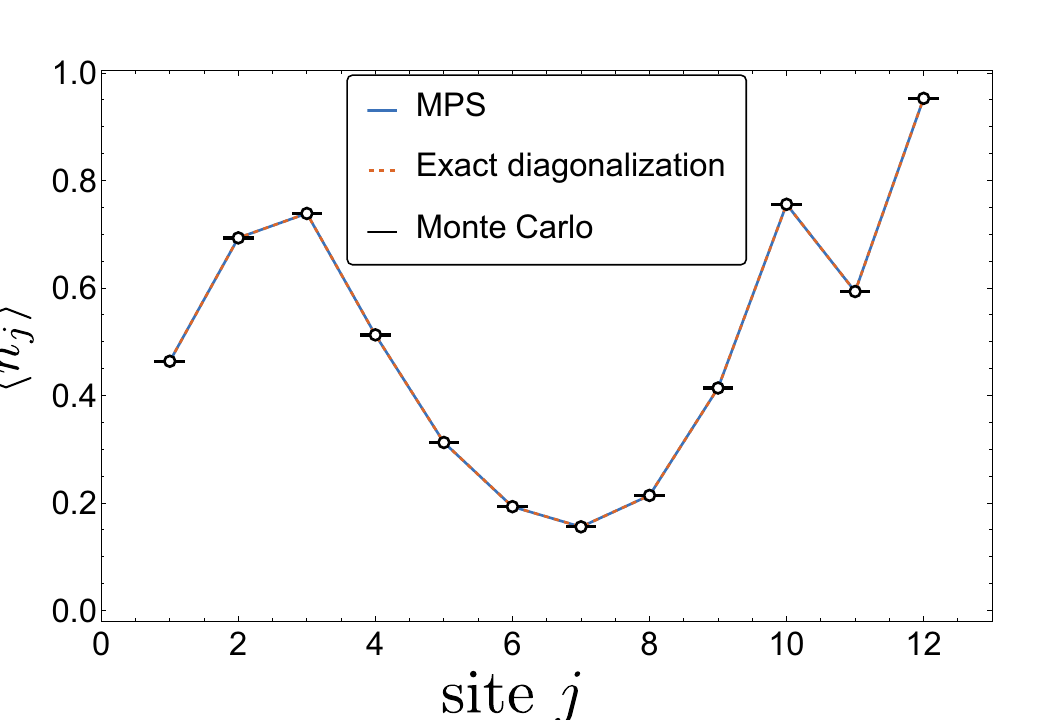}
  \caption{Numerical check of the exact steady state at small system size $L=12$.
  We compare observables computed from the exact matrix-product steady state (MPS) derived in this work, the steady state obtained by exact diagonalization of the Markov generator, $W$, and kinetic Monte Carlo simulations.
  The Monte Carlo data points include error bars (standard errors from finite-sampling statistics).
  The agreement corroborates the correctness of our exact solution.
  \label{Fig-SteadyCheck}}
\end{figure}

To validate the exact steady-state construction at finite size, Fig.~\ref{Fig-SteadyCheck} compares observables from the analytic solution, exact diagonalization, and kinetic Monte Carlo, and shows quantitative agreement.
The exact derivation of this steady state in matrix-product-state (MPS) form is presented later in Sec.~\ref{sec:statistical-mps}.

Throughout this paper, we impose RBC, i.e., a fully closed chain not coupled to particle reservoirs.
Accordingly, no additional boundary terms are added to $W$.
We restrict to the ASEP parameter regime $0<p_{\sigma}<1$ and do not consider the TASEP limits $p_{\sigma}=0,1$.

\subsection{Similarity Transformation to a Hermitian Hamiltonian}\label{subsec:hermitization}

For $0<p_{\sigma}<1$, the non‑Hermitian operator $W$ can be transformed by the invertible but nonunitary operator
\begin{align}
    &S\coloneqq\prod_{\sigma=1}^{M}S_{\sigma}, \quad
    S_{\sigma}\coloneqq q_{\sigma}^{-\sum_{j=1}^{L} j \hat{n}_{j,\sigma}},\nt
    &S^{-1}=\prod_{j=1}^{L}\qty[\qty(1-\hat{n}_{j})
      +\sum_{\sigma=1}^{M}q_{\sigma}^{j}\hat{n}_{j,\sigma}],
\end{align}
into the Hermitian Hamiltonian
\begin{align}
\label{eq:HermHam}
    &H\coloneqq S\,W\,S^{-1}=\sum_{j=1}^{L-1} h_{j,j+1}, \nt
    &h_{j,j+1}=\nt
    &- \sum_{\sigma=1}^{M}\sqrt{p_{\sigma}\qty(1-p_{\sigma})}
      \qty(\hc{b}_{j,\sigma}b_{j+1,\sigma} +\hc{b}_{j+1,\sigma}b_{j,\sigma})\nt
    &+\sum_{\sigma=1}^{M}\qty[p_{\sigma}\qty(1-\hat{n}_{j})\hat{n}_{j+1,\sigma}
      +\qty(1-p_{\sigma})\hat{n}_{j,\sigma}\qty(1-\hat{n}_{j+1})].
\end{align}
This extends the known Hermitianization for single‑species ASEP under RBC~\cite{PhysRevA.46.844}.  
In non‑Hermitian physics, it is known as the Hatano–Nelson pure imaginary gauge transformation with RBC~\cite{PhysRevLett.77.570,PhysRevB.56.8651}.  
Because similarity transformations preserve the spectrum, $H$ has a real spectrum under RBC.  
If $\ket{\Psi}$ is an eigenstate of $H$, then the corresponding probability eigenvector for $W$ is
\begin{align}
    \ket{P}=S^{-1}\ket{\Psi}.
\end{align}
Their relation is illustrated by:
\begin{tikzcd}
  W \arrow[r, "S"] \arrow[d, "\text{eigenstate}"'] 
    & H\coloneqq S\,W\,S^{-1} \arrow[d, "\text{eigenstate}"] \\
  \ket{P}=S^{-1}\ket{\Psi} & \ket{\Psi} \arrow[l, "S^{-1}"].
\end{tikzcd}

\section{Symmetries and  unitary transformations}\label{sec:symmetry}

In this section, we list the symmetries of $W$ and $H$ and discuss symmetries in parameter space.  
We treat the sequence operator $\mathcal{M}$ separately because it is crucial for enumerating exact ground states.

\subsection{Global $\mathrm{U}\qty(1)^{\times M}$ Symmetry}\label{subsec:other-symmetry}

The model has a $\mathrm{U}\qty(1)^{\times M}$ symmetry, which conserves the particle number of each species.
Under independent phase rotations 
$b_{j,\sigma}\mapsto e^{\im\theta_{\sigma}}b_{j,\sigma}\ \qty(\theta_{\sigma}\in\mathbb{R})$,  
both $W$ and $H$ remain invariant.
Hence, each species' particle number 
$\hat{N}_{\sigma}\coloneqq\sum_{j=1}^{L}\hat{n}_{j,\sigma}$ 
is conserved:
\begin{align}
  \com{W}{\hat{N}_{\sigma}}=\com{H}{\hat{N}_{\sigma}}=0.
  \tag{32}
\end{align}
Consequently, the total particle number 
$\hat{N}\coloneqq\sum_{j=1}^{L}\hat{n}_{j}$ 
is conserved as well.

\subsection{Parity and Species‑Swap Unitaries}\label{subsec:mirrorUnitary}

\begin{figure}[H]
  \includegraphics[width=\columnwidth,clip]{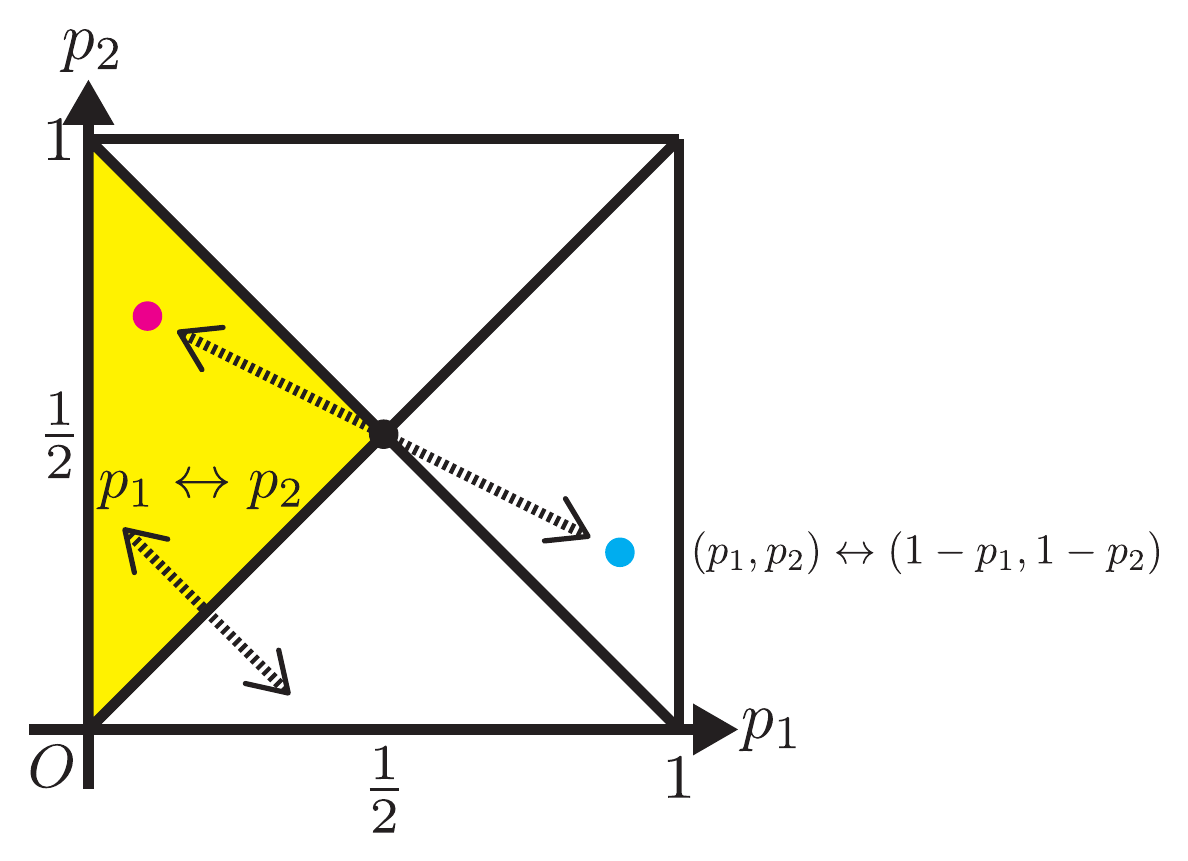}
  \caption{Parameter relations that give the same spectrum under unitary transformations for $M=2$.  
  These relations imply that studying the spectrum in the yellow isosceles triangle is sufficient.  
  \label{Fig-mirror}}
\end{figure}

We define the parity unitary transformation that swaps left and right on the lattice by:
\begin{align}
  \hc{P}\,b_{j,\sigma}\,P = b_{L+1-j,\sigma}.
  \tag{33}
\end{align}
Under this transformation, the spectrum is preserved and the Hamiltonian transforms as
\begin{align}
  \hc{P}\,H\qty(p_{1},\cdots,p_{M})\,P
  &= H\qty(1-p_{1},\cdots,1-p_{M}),
  \tag{34}\label{eq:mirror}
\end{align}
as shown in Fig.~\ref{Fig-mirror}.  
Thus, $H\qty(p_{1},\cdots,p_{M})$ and $H\qty(1-p_{1},\cdots,1-p_{M})$ have the same spectrum.

For $1\le\sigma_{1}<\sigma_{2}\le M$, we define a unitary operator:
\begin{align}
  U_{\sigma_{1},\sigma_{2}}
  &\coloneqq \exp\qty[\im \frac{\pi}{2}
  \sum_{j=1}^{L}\qty(-\im\,\hc{b}_{j,\sigma_{1}}\,b_{j,\sigma_{2}}
  +\im\,\hc{b}_{j,\sigma_{2}}\,b_{j,\sigma_{1}})].
  \tag{35}
\end{align}
This operator acts on the hard-core boson operators as
\begin{align}
  \hc{U}_{\sigma_{1},\sigma_{2}}\,b_{j,\sigma_{1}}\,U_{\sigma_{1},\sigma_{2}}
  &= b_{j,\sigma_{2}},\nt
  \hc{U}_{\sigma_{1},\sigma_{2}}\,b_{j,\sigma_{2}}\,U_{\sigma_{1},\sigma_{2}}
  &= -b_{j,\sigma_{1}},
  \tag{36}\label{eq:12swap}
\end{align}
and it swaps $p_{\sigma_{1}}$ and $p_{\sigma_{2}}$ in the Hamiltonian:
\begin{align}
  &\hc{U}_{\sigma_{1},\sigma_{2}}\,
  H\qty(\cdots,p_{\sigma_{1}},\cdots,p_{\sigma_{2}},\cdots)\,
  U_{\sigma_{1},\sigma_{2}}\nt
  &= H\qty(\cdots,p_{\sigma_{2}},\cdots,p_{\sigma_{1}},\cdots).
  \tag{37}
\end{align}
\eqref{eq:mirror} and \eqref{eq:12swap} imply that it suffices to study the spectrum within the simplex
\begin{align}
  \chuukakko{
  \qty(p_{1},\cdots,p_{M})\in\qty(0,1)^{\times M}
  \mid p_{1}\le\cdots\le p_{M},\ p_{1}+p_{M}\le1}.
  \tag{38}
\end{align}
In the following sections, when we treat the concrete examples, we focus on the $M=2$ case and restrict our analysis to the yellow region in Fig.~\ref{Fig-mirror}.

\subsection{Particle Sequence Conservation and Hilbert Space Fragmentation}\label{subsec:sequence-preserving}

The continuous time Markov dynamics generated by $W$ forbids particle overtaking.  
Therefore, the sequence of particles 
$S=(\sigma_{1},\cdots,\sigma_{N})$ 
is preserved.  
To formalize this, we introduce the sequence operator:
\begin{align}
  \mathcal{M}
  &\coloneqq \sum_{j=1}^{L}M^{\hat{n}_{1}+\cdots+\hat{n}_{j-1}}
  \sum_{\sigma=1}^{M}\sigma\,\hat{n}_{j,\sigma}.
  \tag{39}\label{eq:sequenceop}
\end{align}
This operator commutes with both $H$ and $W$:
\begin{align}
  \com{H}{\mathcal{M}}=\com{W}{\mathcal{M}}=0.
  \tag{40}
\end{align}
It has $\sum_{k=0}^{L}M^{k}=\frac{M^{L+1}-1}{M-1}$ distinct eigenvalues.  
Each configuration $S$ maps to a unique base‑$M$ eigenvalue.  
For example, for $M=2$ and $N=0,1,2$, the eigenvalues are:
\begin{align}
  &0: \ket{\Omega}, \nt
  &1: \hc{b}_{j,1}\ket{\Omega}\quad\qty(1\le j\le L), \nt
  &2: \hc{b}_{j,2}\ket{\Omega}\quad\qty(1\le j\le L), \nt
  &3: \hc{b}_{j_{1},1}\hc{b}_{j_{2},1}\ket{\Omega}\quad\qty(1\le j_{1}<j_{2}\le L), \nt
  &4: \hc{b}_{j_{1},2}\hc{b}_{j_{2},1}\ket{\Omega}\quad\qty(1\le j_{1}<j_{2}\le L), \nt
  &5: \hc{b}_{j_{1},1}\hc{b}_{j_{2},2}\ket{\Omega}\quad\qty(1\le j_{1}<j_{2}\le L), \nt
  &6: \hc{b}_{j_{1},2}\hc{b}_{j_{2},2}\ket{\Omega}\quad\qty(1\le j_{1}<j_{2}\le L).
  \tag{41}
\end{align}
Hence, $H$ is block diagonal in $\mathcal{M}$:
\begin{align}
  H=\bigoplus_{S}H^{S}
  =\bigoplus_{m=0}^{\frac{M^{L+1}-1}{M-1}-1}H^{\mathcal{M}=m}.
  \tag{42}
\end{align}
\addtocounter{equation}{11}
Here, $H^{S}$ and $H^{\mathcal{M}=m}$ are the blocks corresponding to the particle species sequence $S$ and the eigenvalue $m$, respectively.  
This fragmentation into exponentially many sectors leads to Hilbert space fragmentation (HSF) and consequently violates the eigenstate thermalization hypothesis (ETH)\cite{PhysRevX.10.011047,Moudgalya_2022,miura_future}.  
This type of HSF due to forbidden overtaking resembles that in the Hubbard model with infinite on-site repulsion~\cite{PhysRevB.41.2326,PhysRevLett.134.010411} and the $t$-$J_{z}$ model \cite{PhysRevB.101.125126}.

\begin{figure}[H]
  \includegraphics[width=\columnwidth,clip]{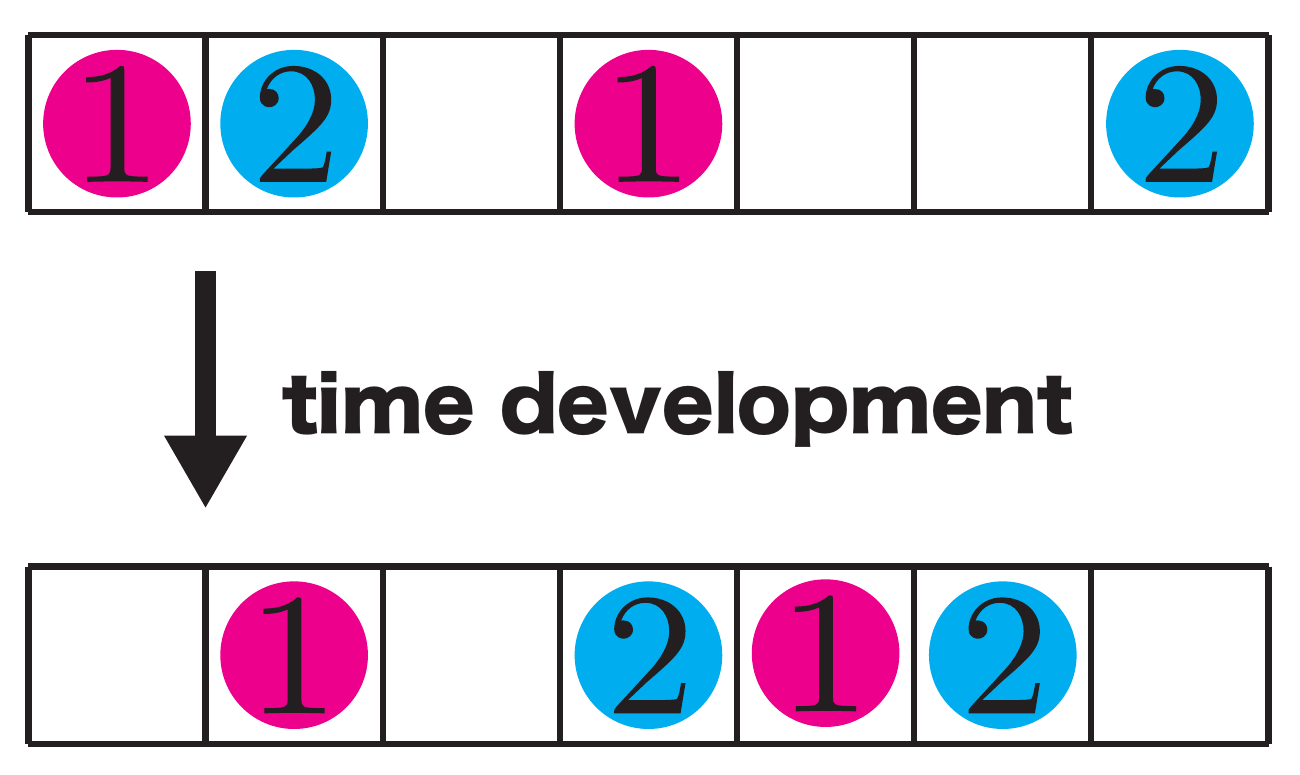}
  \caption{Example time evolution for $S=(1,2,1,2)$.  
  Particles can only move without changing the particle species sequence $(1,2,1,2)$.  
  The motion is highly constrained in the full $3^L$-dimensional Hilbert space.  
  \label{Fig-NoOvertaking}}
\end{figure}

In Sec.~\ref{sec:ground_state}, we show from the Markov process structure that each of the $\frac{M^{L+1}-1}{M-1}$ sectors has a unique ground state.  
We then show that these ground states exhaust all ground states of $H$ and are exactly given by the form of matrix product state (MPS).  
Although similar to the matrix product ansatz reviewed in \cite{Blythe_2007}, here the bond dimension is exactly $N+1$ for a system with $N$ particles and remains finite for finite $N$.
A key structural difference from the boundary-driven matrix-product ansatz in Ref.~\cite{Blythe_2007} is that our auxiliary index tracks the ordered particle species sequence sector fixed by the no-passing constraint: the off-diagonal auxiliary step inserts the next species in that fixed order rather than selecting a site-dependent algebra element by boundary injection or extraction rules.
This is why the present construction is directly suited to sector-resolved steady observables under reflecting (closed) boundary conditions.

\subsection{Enhanced Symmetry in the Isotropic Limit}

When $p_{1}=\cdots=p_{M}$, $H$ acquires an additional high symmetry.  
It commutes with the local operators $\hc{b}_{j,\sigma}b_{j,\tau}$ for any $j$ and any $\sigma\neq\tau$.  
This reflects that, under isotropic hopping, particle species become indistinguishable.  
Any eigenstate that contains particles of more than one species can be converted into some eigenstates containing only a single particle species without changing its energy by applying $\hc{b}_{j,\sigma}b_{j,\tau}$.  
Combined with the discussion in Appendix.~\ref{app:sector}, this implies the spectrum (up to degeneracy) matches that of the standard ASEP.

\section{Exact steady state and exponential computational complexity}\label{sec:exact-steady-complexity}

Since the system considered in this paper is finite, a probability distribution $\pi_{\mathrm{s.s.}}(C)$ that satisfies the detailed-balance condition
\begin{align}
  W_{C'\leftarrow C}\pi_{\mathrm{s.s.}}(C)
  =W_{C\leftarrow C'}\pi_{\mathrm{s.s.}}(C')
\end{align}
is a steady state.
In this section, we construct the steady-state weights in each fixed particle species sequence sector from detailed balance.
We then explain why the later MPS-transfer-matrix formulation is still essential in practice: although detailed balance gives explicit weights, direct observable evaluation requires summing over all allowed configurations and is exponentially costly in the worst case.
By contrast, the MPS-transfer-matrix approach yields closed-form expressions for one- and two-point observables with polynomial complexity.

\subsection{Exact steady states for each particle species sector}

Fix one particle species sector $S=(\sigma_{1},\ldots,\sigma_{N})$.
Then the stationary distribution $\pi_{\mathrm{s.s.}}(C)$ in sector $S$ is given, with particle positions $j_{1},\ldots,j_{N}$ ($j_{1}<\cdots<j_{N}$), by
\begin{align}
  \pi_{\mathrm{s.s.}}(C)
  \propto \prod_{n=1}^{N}q_{\sigma_{n}}^{2j_{n}}
  =\prod_{n=1}^{N}\qty(\frac{1-p_{\sigma_{n}}}{p_{\sigma_{n}}})^{j_{n}},
\end{align}
and one can show that it satisfies detailed balance.

The proof is as follows.
Since the particle species of each particle are fixed by $S$, configurations $C,C'$ can be specified only by the position data of each particle.
Now let $j_{n+1}\neq j_{n}+1$, and set
\begin{align}
  C&=(j_{1},\ldots,j_{n},\ldots,j_{N}),\\
  C'&=(j_{1},\ldots,j_{n}+1,\ldots,j_{N}).
\end{align}
Between these two configurations, there exist the transition $C\to C'$ by a right jump of the $n$th particle and the transition $C'\to C$ by a left jump.
In this case, the corresponding off-diagonal matrix elements are
\begin{align}
  W_{C'\leftarrow C}&=-(1-p_{\sigma_{n}}),\\
  W_{C\leftarrow C'}&=-p_{\sigma_{n}}.
\end{align}
Moreover,
\begin{align}
  \frac{\pi_{\mathrm{s.s.}}(C')}{\pi_{\mathrm{s.s.}}(C)}
  =\frac{\prod_{n'=1}^{N}\qty(\frac{1-p_{\sigma_{n'}}}{p_{\sigma_{n'}}})^{j_{n'}+\delta_{n,n'}}}
  {\prod_{n'=1}^{N}\qty(\frac{1-p_{\sigma_{n'}}}{p_{\sigma_{n'}}})^{j_{n'}}}
  =\frac{1-p_{\sigma_{n}}}{p_{\sigma_{n}}}.
\end{align}
Therefore,
\begin{align}
  W_{C'\leftarrow C}\pi_{\mathrm{s.s.}}(C)
  &=-(1-p_{\sigma_{n}})\pi_{\mathrm{s.s.}}(C)\nt
  &=-p_{\sigma_{n}}\pi_{\mathrm{s.s.}}(C')\nt
  &=W_{C\leftarrow C'}\pi_{\mathrm{s.s.}}(C'),
\end{align}
so detailed balance holds.

\subsection{Computational complexity for calculating expectation values}\label{subsec:complexity}

In probability-vector notation, the steady-state vector $\ket{P_{\mathrm{s.s.}}}$ corresponding to $\pi_{\mathrm{s.s.}}(C)$ is given by
\begin{align}
  \ket{P_{\mathrm{s.s.}}}
  \propto
  \sum_{1\le j_{1}<\cdots<j_{N}\le L}
  q_{\sigma_{1}}^{2j_{1}}\cdots q_{\sigma_{N}}^{2j_{N}}
  \hc{b}_{j_{1},\sigma_{1}}\cdots\hc{b}_{j_{N},\sigma_{N}}\ket{\Omega}.
\end{align}
As a forward preview, one may write the same steady state in the MPS form that will be derived later as
\begin{align}
  \ket{P_{\mathrm{s.s.}}}\propto \bmm{e}_{1}^{\top}\mathsf{B}_{1}\cdots\mathsf{B}_{L}\bmm{e}_{N+1},
\end{align}
where the local tensors $\mathsf{B}_{j}$ are defined in Sec.~\ref{sec:statistical-mps}.
If one takes this MPS form as an ad hoc ansatz and expands it in the real-space basis, then one indeed recovers the detailed-balance weights above.
However, a more natural route to this MPS structure is available: it is obtained systematically from the ground-state MPS of the Hermitian partner and the similarity transformation, as explained in Secs.~\ref{sec:ground_state} and \ref{sec:statistical-mps}.

To calculate expectation values directly from the configuration-space form, one must explicitly scan all configurations satisfying $1\le j_{1}<\cdots<j_{N}\le L$.
Since the number of configurations is $\binom{L}{N}$, the computational cost is at least $O\qty(\binom{L}{N})$.
In particular, $\binom{L}{N}$ is maximized at $N\simeq L/2$, and
\begin{align}
  \binom{L}{L/2}\sim \frac{2^{L}}{\sqrt{\pi L/2}},
\end{align}
so the worst-case complexity is exponential, $\sim O(L^{-1/2}2^{L})$.
By contrast, in the matrix-product-state representation derived below, transfer-matrix contractions scale as $O(LN^{2})$.
At $N\simeq L/2$, this becomes $O(L^{3})$.
Hence, observable evaluation is reduced from exponential to polynomial complexity.

\section{Exact ground states of ``Quantum" Hamiltonian $H$}\label{sec:ground_state}

In this section, we treat the Hermitian Hamiltonian \eqref{eq:HermHam} as a quantum system.
The similarity transformation $S$ gives a one-to-one correspondence between the spectra of $W$ and $H$, and steady states of $W$ are obtained by applying $S^{-1}$ to ground states of $H$ (Sec.~\ref{subsec:hermitization}).
Moreover, $H$ is frustration free, so its ground states are naturally analyzed with one-dimensional MPS methods.
We determine the number of ground states and derive their explicit forms.
Because each sector of $W$ is a finite irreducible continuous-time Markov process under RBC, each sector has a unique steady state.
Via the similarity transformation, each sector labeled by the eigenvalue of the particle species sequence operator $\mathcal{M}$ therefore has a unique ground state of $H$.
Moreover, each ground state admits the MPS representation with the bond dimension equal to the particle number plus one.

\subsection{Ground-state uniqueness in each particle species sequence sector}\label{subsec:uniqueness}

Let $H^{\mathcal{M}=m}$ be the restriction of $H$ to the sector with $\mathcal{M}=m$ ($m=0,1,\cdots,\frac{M^{L+1}-1}{M-1}-1$).
For fixed $m$, the corresponding block $W^{\mathcal{M}=m}$ acts on a finite configuration space with fixed particle number and fixed particle species sequence.
Under $0<p_{\sigma}<1$, any two configurations in this sector are connected by allowed local hops, so the sector dynamics are irreducible.
Hence the continuous-time Markov process generated by $W^{\mathcal{M}=m}$ has a unique stationary vector.
Since $H^{\mathcal{M}=m}=S W^{\mathcal{M}=m}S^{-1}$, the zero-energy state of each $H^{\mathcal{M}=m}$ is also unique.
Therefore the full Hamiltonian has $\frac{M^{L+1}-1}{M-1}$ ground states in total, one per sector.

\subsection{Local Projector Form and Frustration‑Free Nature}

We define the normalized two‑site state
\begin{align}
    \ket{Q_{-}\qty(\sigma)}
    &\coloneqq \sqrt{1-p_{\sigma}}\ket{\sigma,0}
      -\sqrt{p_{\sigma}}\ket{0,\sigma}\nt
    &=\sqrt{p_{\sigma}}\qty(q_{\sigma}\ket{\sigma,0}-\ket{0,\sigma}).
\end{align}
Then each two‑site term of the Hamiltonian \eqref{eq:HermHam} can be written as
\begin{align}
    h_{j,j+1}
    =\sum_{\sigma=1}^{M}\ket{Q_{-}\qty(\sigma)}\bra{Q_{-}\qty(\sigma)}_{j,j+1}.
\end{align}
Hence $H$ is a sum of projectors.
The product ground states $\ket{\Omega}$ and $\ket{\sigma_{1},\cdots,\sigma_{L}}$ confirm that $H$ is frustration‑free.

We define
\begin{align}
    \ket{Q_{+}\qty(\sigma)}
    &\coloneqq \sqrt{p_{\sigma}}\ket{\sigma,0}
      +\sqrt{1-p_{\sigma}}\ket{0,\sigma}\nt
    &=\sqrt{p_{\sigma}}\qty(\ket{\sigma,0}+q_{\sigma}\ket{0,\sigma}).
\end{align}
Then, the eigensystem of $h_{j,j+1}$ is:
\begin{align}
    E&=0:\nt
    &\ket{0,0},\nt
    &\ket{Q_{+}\qty(\sigma)}\quad\qty(\sigma=1,\cdots,M),\nt
    &\ket{\sigma_{1},\sigma_{2}}\quad\qty(\sigma_{1},\sigma_{2}=1,\cdots,M),\nt
    E&=1:\nt
    &\ket{Q_{-}\qty(\sigma)}\quad\qty(\sigma=1,\cdots,M),
\end{align}
in total $\qty(M+1)^{2}$ states.  
From this, zero‑energy ground states of the full system can be built as MPS.

\subsection{Exact MPS Representation of ground states}\label{subsec:ssMPS}

Let $N$ be the particle number.  
On each site, define $\qty(N+1)\times\qty(N+1)$ matrices $\mathsf{A}$.
Before that, we define the following:
\begin{align}
\label{eq:lambdas}
    \lambda_{m} &\coloneqq
\begin{cases}
\qty(q_{\sigma_m}\cdots q_{\sigma_N})^2, & m=1,\cdots,N,\\
1, & m=N+1.
\end{cases}
\end{align}
Then we set
\begin{align}
 &\mathsf{A} \coloneqq
\begin{pmatrix}
\sqrt{\lambda_{1}}\ket{0}      & \ket{\sigma_{1}} &               &               &               \\
                 & \sqrt{\lambda_{2}}\ket{0}   & \ket{\sigma_{2}} &            &               \\
                 &                  & \ddots        & \ddots        &               \\
                 &                  &               & \sqrt{\lambda_{N}}\ket{0}  & \ket{\sigma_{N}} \\
                 &                  &               &               & \sqrt{\lambda_{N+1}}\ket{0}
\end{pmatrix}.
\end{align}
We also define
\begin{align}
\ket{\bar{Q}_{+}\qty(\sigma)}\coloneqq\ket{\sigma,0}+q_{\sigma}\ket{0,\sigma}.
\end{align}
Then for two sites, we have:
\begin{widetext}
    \begin{align}
    &\mathsf{A}_{j}\mathsf{A}_{j+1}
    =
    \begin{pmatrix}
        \lambda_{1}\ket{0,0}           & \sqrt{\lambda_{2}}\ket{\bar{Q}_{+}\qty(\sigma_{1})}       & \ket{\sigma_{1},\sigma_{2}}    &               &               &               \\
                                 & \lambda_{2}\ket{0,0}             & \sqrt{\lambda_{3}}\ket{\bar{Q}_{+}\qty(\sigma_{2})}       & \ket{\sigma_{2},\sigma_{3}}    &               &               \\
                                 &                                & \ddots                        & \ddots                          & \ddots        &               \\
                                 &                                &                                & \lambda_{N-1}\ket{0,0}                 & \sqrt{\lambda_{N}}\ket{\bar{Q}_{+}\qty(\sigma_{N-1})}       & \ket{\sigma_{N-1},\sigma_{N}} \\
                                 &                                &                                &                                & \lambda_{N}\ket{0,0}                & \sqrt{\lambda_{N+1}}\ket{\bar{Q}_{+}\qty(\sigma_{N})}         \\
                                 &                                &                                &                                &                                & \lambda_{N+1}\ket{0,0}                
    \end{pmatrix}.
    \end{align}
\end{widetext}
Since each entry is in the kernel of $h_{j,j+1}$,
\begin{align}
    h_{j,j+1}\,\mathsf{A}_{j}\mathsf{A}_{j+1}
    =\raisebox{-5.0ex}{\includegraphics[scale=0.3]{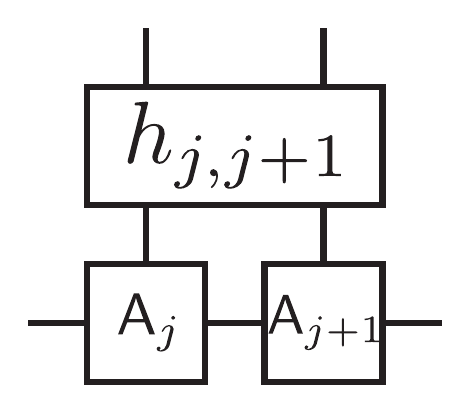}}=0.
\end{align}
We denote by $\bmm{e}_{m}$ as the unit vector whose 
$m$‑th component is $1$ and all other components are $0$.
Choosing boundary vectors $\bmm{e}_{1}^{\top}$ and $\bmm{e}_{N+1}$ for the reflecting (closed) chain MPS,
\begin{align}
\label{eq:Psidash}
    &\ket{\Psi'_{\text{g.s.}}}\nt
    =&
    \raisebox{-3.0ex}{\includegraphics[scale=0.3]{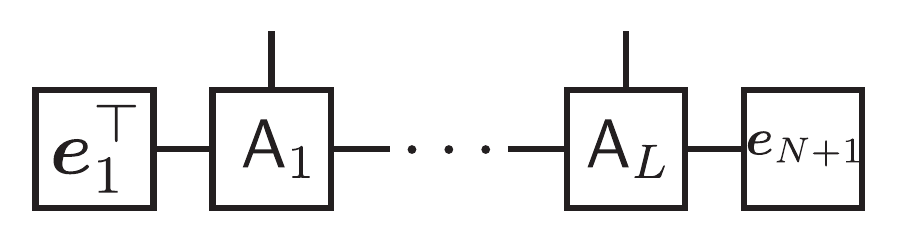}}\nt
    =&\bmm{e}_{1}^{\top}\mathsf{A}_{1}\cdots\mathsf{A}_{L}\bmm{e}_{N+1}\nt
    =&\sum_{1\leq j_{1}<\cdots<j_{N}\leq L}
    q^{j_{1}-1}_{\sigma_{1}}\cdots q^{j_{N}-N}_{\sigma_{N}}
    \hc{b}_{j_{1},\sigma_{1}}\cdots\hc{b}_{j_{N},\sigma_{N}}
    \ket{\Omega}.
\end{align}
Note that $\ket{\Psi'_{\text{g.s.}}}$ is not normalized and these ground states include the product states $\ket{\Omega}$ and $\ket{\sigma_{1},\cdots,\sigma_{L}}$.
There are $\sum_{N=0}^{L}\qty(\text{number of choise of }q_{\sigma})=\sum_{N=0}^{L}M^{N}=\frac{M^{L+1}-1}{M-1}$ such states, one in each particle species sequence sector.  

These ground states can also be constructed algebraically.  
See Appendix~\ref{app:algebraic}.

\section{Exact steady state of the transition matrix $W$}\label{sec:statistical-mps}

In Sec.~\ref{sec:ground_state}, we obtained ground states $\ket{\Psi'_{\text{g.s.}}}$ of the Hermitian Hamiltonian $H$.
Here we construct steady states of $W$ by applying $S^{-1}$ and then normalizing.
This gives closed-form formulas for steady one- and two-point observables.
For $M=2$, we evaluate the density and correlation profiles and compare them with standard ASEP behavior.
As emphasized in Sec.~\ref{sec:exact-steady-complexity}, the MPS form is also computationally advantageous.
Direct evaluation from detailed balance requires a sum over $\binom{L}{N}$ configurations (worst-case exponential cost), whereas transfer-matrix contractions in the MPS form scale as $O(LN^{2})$.

\subsection{Constructing the Steady State MPS}

The unnormalized steady vector is
\begin{align}
  &\ket{P'_{\text{s.s.}}}\nt
  \coloneqq &S^{-1}\ket{\Psi'_{\text{g.s.}}}\nt
  =&S^{-1}\times \raisebox{-3.0ex}{\includegraphics[scale=0.3]{fig07.pdf}}\nt
  =& \sum_{1\le j_{1}<\cdots<j_{N}\le L}
     q^{2j_{1}-1}_{\sigma_{1}}\cdots q^{2j_{N}-N}_{\sigma_{N}}
     \hc{b}_{j_{1},\sigma_{1}}\cdots\hc{b}_{j_{N},\sigma_{N}}
     \ket{\Omega}\nt
  =& \bmm{e}_{1}^{\top}\mathsf{B}_{1}\cdots\mathsf{B}_{L}\bmm{e}_{N+1}\nt
  =& \raisebox{-2.0ex}{\includegraphics[scale=0.3]{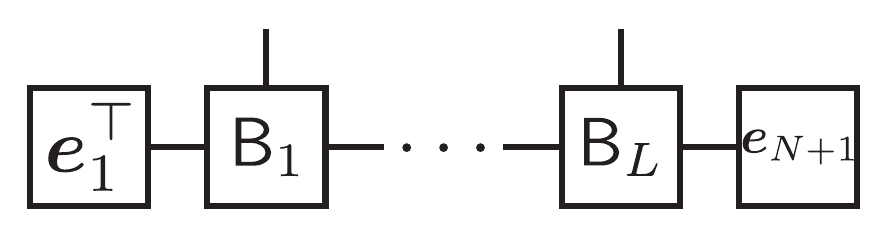}}.
\end{align}
The powers here differ from the detailed-balance weights in Sec.~\ref{sec:exact-steady-complexity} only by a position-independent constant within a fixed particle species sequence sector, and therefore define the same normalized steady state.
Here, $\mathsf{B}$ is defined by
\begin{align}
  \mathsf{B}
  &\coloneqq
\begin{pmatrix}
\lambda_{1}\ket{0}      & \ket{\sigma_{1}} &               &               &               \\
                        & \lambda_{2}\ket{0}   & \ket{\sigma_{2}} &            &               \\
                        &                  & \ddots        & \ddots        &               \\
                        &                  &               & \lambda_{N}\ket{0}  & \ket{\sigma_{N}} \\
                        &                  &               &               & \lambda_{N+1}\ket{0}
\end{pmatrix}.
\end{align}
For each fixed particle species sequence $S$, these states are linearly independent because the eigenvalue of $\mathcal{M}$ is different.  
Each sector with $N$ particles has $M^{N}$ independent states.  
Hence the total number of unnormalized steady states is the same as $\ket{P'_{\text{s.s.}}}$,
$\frac{M^{L+1}-1}{M-1}$.
We compute the normalization constant in Sec.~\ref{subsec:statistical-mps-normalize}.

\subsection{Normalization of the MPS Steady State}\label{subsec:statistical-mps-normalize}

The normalization constant of the steady state $\ket{P'_{\text{s.s.}}}$ is
\begin{align}
  \mathcal{N}
  &\coloneqq \sum_{\{\mu\}}\braket{\{\mu\}}{P'_{\text{s.s.}}}.
\end{align}
The stochastic transfer matrix is defined as
\begin{align}
\label{eq:TransferB}
    \Tilde{\mathsf{B}}&\coloneqq\raisebox{-5.0ex}{\includegraphics[scale=0.3]{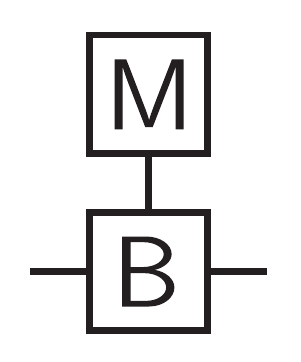}}=\sum_{\mu}\bra{\mu}\mathsf{B}\nt
    &=
    \begin{pmatrix}
\lambda_{1}      & 1 &               &               &               \\
                 & \lambda_{2}   & 1 &            &               \\
                 &                  & \ddots        & \ddots        &               \\
                 &                  &               & \lambda_{N}  & 1 \\
                 &                  &               &               & \lambda_{N+1}
\end{pmatrix},
\end{align}
where $\mathsf{M}\coloneqq\sum_{\mu=0}^{M}\bra{\mu}$.
Then
\begin{align}
    \mathcal{N}&=\raisebox{-5.0ex}{\includegraphics[scale=0.3]{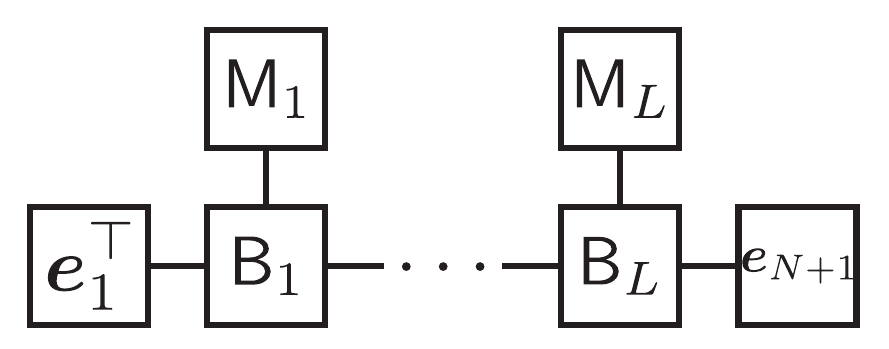}}\nt
    &=\bmm{e}_{1}^{\top}\Tilde{\mathsf{B}}^{L}\bmm{e}_{N+1}\nt
    &=\qty(\Tilde{\mathsf{B}}^{L})_{1,N+1}.
\end{align}
By induction one shows for $\forall n\in \mathbb{Z}_{\geq 0}$,
\begin{align}
\label{eq:Bmatpower}
  \qty(\Tilde{\mathsf{B}}^{n})_{i,j}
  = \begin{cases}
    0, & i>j,\\
    \displaystyle
    \sum_{l=i}^{j}
    \frac{\lambda_{l}^{\,n}}
         {\prod_{\substack{i\le l'\le j\\l'\neq l}}\qty(\lambda_{l}-\lambda_{l'})},
    & i\le j.
  \end{cases}
\end{align}
Hence
\begin{align}
  \mathcal{N}
  &= \sum_{l=1}^{N+1}
    \frac{\lambda_{l}^{\,L}}
         {\prod_{\substack{1\le l'\le N+1\\l'\neq l}}\qty(\lambda_{l}-\lambda_{l'})}.
\end{align}
If two eigenvalues coincide, one has to take the appropriate limit in this formula.  
The normalized steady state is
\begin{align}
\label{eq:Pss}
    &\ket{P_{\text{s.s.}}}\nt
    =&\frac{\bmm{e}_{1}^{\top}\mathsf{B}_{1}\cdots\mathsf{B}_{L}\bmm{e}_{N+1}}{\mathcal{N}}
    \nt
    =&\frac{\raisebox{-5.0ex}{\includegraphics[scale=0.3]{fig08.pdf}}}{\raisebox{-5.0ex}{\includegraphics[scale=0.3]{fig10.pdf}}}\nt
    =&\frac{\sum\limits_{1\leq j_{1}<\cdots<j_{N}\leq L}
    q^{2j_{1}-1}_{\sigma_{1}}\cdots q^{2j_{N}-N}_{\sigma_{N}}
    \hc{b}_{j_{1},\sigma_{1}}\cdots\hc{b}_{j_{N},\sigma_{N}}
    \ket{\Omega}}
    {\sum_{l=1}^{N+1} \frac{\lambda_{l}^{L}}{\prod\limits_{\substack{1\leq l' \leq N+1 \\l' \neq l}} (\lambda_{l} - \lambda_{l'})}}.
\end{align}

\subsection{Particle distribution and correlation via transfer-matrix method}\label{subsec:distribution-correlation}

The stochastic expectation of an observable $A$ is
\begin{align}
  \kitaiti{A}
  = \sum_{\{C\}} \bra{\{C\}}A\ket{P_{\text{s.s.}}}.
\end{align}
If $A$ is a matrix product operator (MPO), one can use the transfer-matrix method to compute $\kitaiti{A}$.
The explicit formulas written in terms of $\lambda_l$ are exact, algebraically organized results.
For practical numerical evaluation, however, we directly contract the transfer-matrix products in Eqs.~\eqref{eq:numdens} and \eqref{eq:numcorr}; this scales as $O(LN^{2})$.
Compared with the direct detailed-balance configuration summation discussed in Sec.~\ref{subsec:complexity}, the MPS-transfer-matrix evaluation is far more efficient.

\subsubsection{Exact expression for $\kitaiti{\hat{n}_{j}}$}

We define
\begin{align}
\label{eq:TransferN}
    \Tilde{\mathsf{N}}&\coloneqq
    \raisebox{-8.0ex}{\includegraphics[scale=0.3]{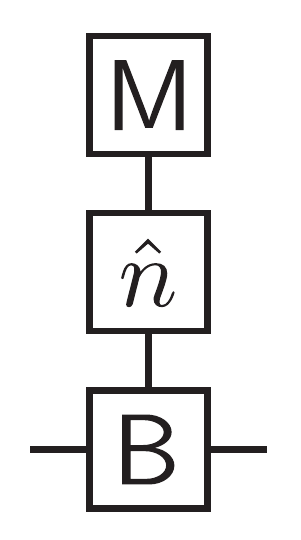}}=\sum_{\mu}\bra{\mu}\hat{n}\mathsf{B}\nt
    &=
    \begin{pmatrix}
0       & 1 &               &               &               \\
                 & 0   & 1 &            &               \\
                 &                  & \ddots        & \ddots        &               \\
                 &                  &               & 0  & 1 \\
                 &                  &               &               & 0
\end{pmatrix}.
\end{align}
Then the density at site $j$ is
\begin{align}
\label{eq:numdens}
    \kitaiti{\hat{n}_{j}}=&\sum_{\{C\}}\bra{\{C\}}\hat{n}_{j}\ket{P_{\text{s.s.}}}\nt
    =&\frac{\raisebox{-5.0ex}{\includegraphics[scale=0.3]{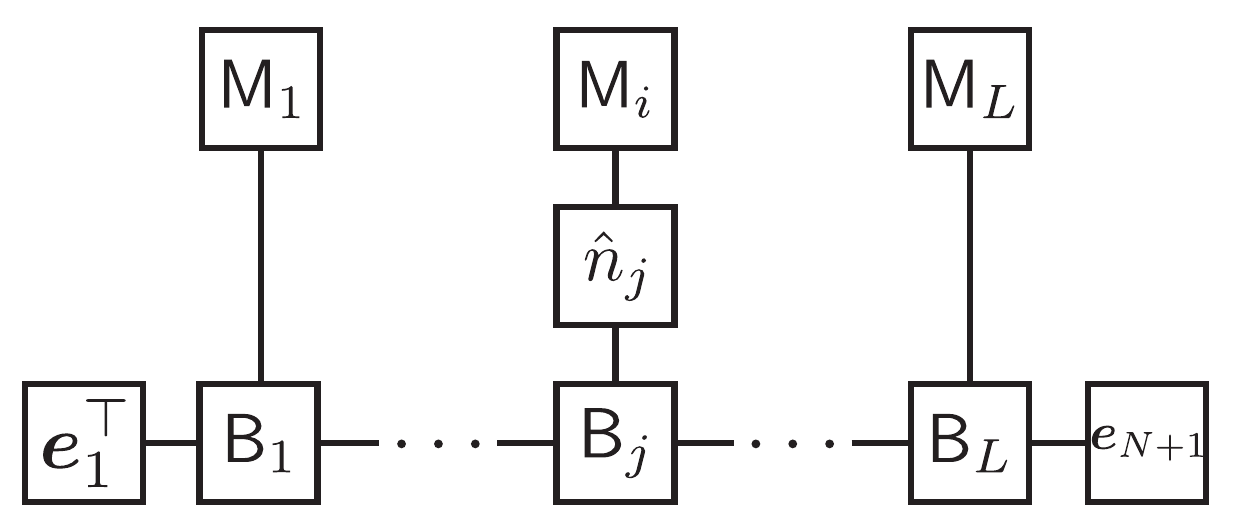}}}{\raisebox{-5.0ex}{\includegraphics[scale=0.3]{fig10.pdf}}}\nt
    =&\frac{\bmm{e}_{1}^{\top}\Tilde{\mathsf{B}}^{j-1}\Tilde{\mathsf{N}}\Tilde{\mathsf{B}}^{L-j}\bmm{e}_{N+1}}{\mathcal{N}}.
\end{align}
Since
\begin{align}
  \Tilde{\mathsf{N}}
  = \sum_{m=1}^{N} \bmm{e}_{m}\bmm{e}_{m+1}^{\top},
\end{align}
we get
\begin{align}
    \kitaiti{\hat{n}_{j}}&=\sum_{m=1}^{N}
    \frac{\bmm{e}_{1}^{\top}\Tilde{\mathsf{B}}^{j-1}\bmm{e}_{m}\bmm{e}_{m+1}^{\top}\Tilde{\mathsf{B}}^{L-j}\bmm{e}_{N+1}}{\mathcal{N}}\nt
    &=\sum_{m=1}^{N}
    \frac{\qty(\Tilde{\mathsf{B}}^{j-1})_{1,m}\qty(\Tilde{\mathsf{B}}^{L-j})_{m+1,N+1}}{\qty(\Tilde{\mathsf{B}}^{L})_{1,N+1}}.
\end{align}
Using \eqref{eq:Bmatpower}, this becomes
\begin{widetext}
    \begin{align}
  \kitaiti{\hat{n}_{j}}
  = \sum_{m=1}^{N}
    \frac{\qty(\sum_{l=1}^{m}\frac{\lambda_{l}^{\,j-1}}
                                 {\prod_{\substack{1\le l'\le m\\l'\neq l}}\qty(\lambda_{l}-\lambda_{l'})})
          \qty(\sum_{l=m+1}^{N+1}\frac{\lambda_{l}^{\,L-j}}
                                    {\prod_{\substack{m+1\le l'\le N+1\\l'\neq l}}\qty(\lambda_{l}-\lambda_{l'})})}
         {\sum_{l=1}^{N+1}\frac{\lambda_{l}^{\,L}}
                            {\prod_{\substack{1\le l'\le N+1\\l'\neq l}}\qty(\lambda_{l}-\lambda_{l'})}}.
\end{align}

\subsubsection{Exact expression for $\kitaiti{\hat{n}_{i}\hat{n}_{j}}$}\label{subsec:numcorr}
The two‑point correlation function $\kitaiti{\hat{n}_{i}\hat{n}_{j}}$ for $i<j$ can be expressed using the MPS transfer matrix as
\begin{align}
    \label{eq:numcorr}
    \kitaiti{\hat{n}_{i}\hat{n}_{j}}
    &=\sum_{\{C\}}\bra{\{C\}}\hat{n}_{i}\hat{n}_{j}\ket{P_{\text{s.s.}}}\nt
    &=\frac{\raisebox{-5.0ex}{\includegraphics[scale=0.3]{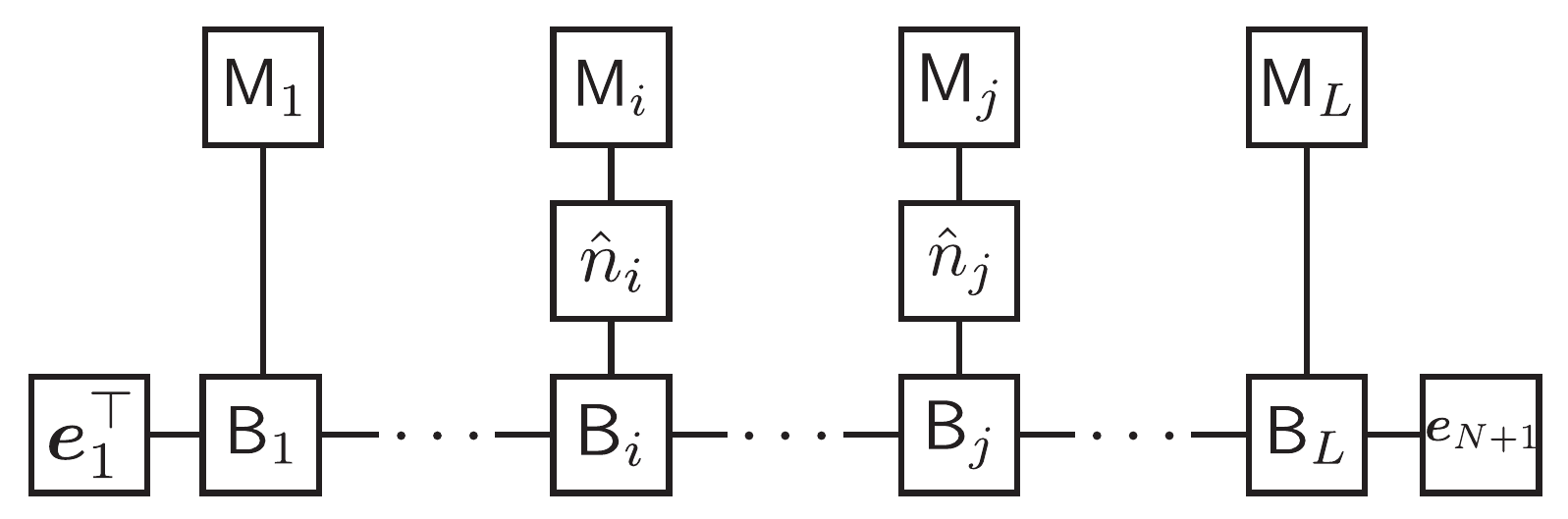}}}
            {\raisebox{-5.0ex}{\includegraphics[scale=0.3]{fig10.pdf}}}\nt
    &=\frac{\bmm{e}_{1}^{\top}\,\Tilde{\mathsf{B}}^{\,i-1}\,\Tilde{\mathsf{N}}\,
             \Tilde{\mathsf{B}}^{\,j-i-1}\,\Tilde{\mathsf{N}}\,
             \Tilde{\mathsf{B}}^{\,L-j}\,\bmm{e}_{N+1}}{\mathcal{N}}\nt
    &=\sum_{m_{i},m_{j}=1}^{N}
    \frac{\bmm{e}_{1}^{\top}\,\Tilde{\mathsf{B}}^{\,i-1}\,\bmm{e}_{m_{i}}\,
          \bmm{e}_{m_{i}+1}^{\top}\,\Tilde{\mathsf{B}}^{\,j-i-1}\,\bmm{e}_{m_{j}}\,
          \bmm{e}_{m_{j}+1}^{\top}\,\Tilde{\mathsf{B}}^{\,L-j}\,\bmm{e}_{N+1}}
         {\mathcal{N}}\nt
    &=\sum_{m_{i},m_{j}=1}^{N}
    \frac{\qty(\Tilde{\mathsf{B}}^{\,i-1})_{1,m_{i}}\,
          \qty(\Tilde{\mathsf{B}}^{\,j-i-1})_{m_{i}+1,m_{j}}\,
          \qty(\Tilde{\mathsf{B}}^{\,L-j})_{m_{j}+1,N+1}}
         {\qty(\Tilde{\mathsf{B}}^{\,L})_{1,N+1}}\nt
    &=\sum_{1\le m_{i}<m_{j}\le N}
    \frac{\displaystyle
      \qty(\sum_{l=1}^{m_{i}}
        \frac{\lambda_{l}^{\,i-1}}
             {\prod_{\substack{1\le l'\le m_{i}\\l'\neq l}}\qty(\lambda_{l}-\lambda_{l'})})\,
      \qty(\sum_{l=m_{i}+1}^{m_{j}}
        \frac{\lambda_{l}^{\,j-i-1}}
             {\prod_{\substack{m_{i}+1\le l'\le m_{j}\\l'\neq l}}\qty(\lambda_{l}-\lambda_{l'})})\,
      \qty(\sum_{l=m_{j}+1}^{N+1}
        \frac{\lambda_{l}^{\,L-j}}
             {\prod_{\substack{m_{j}+1\le l'\le N+1\\l'\neq l}}\qty(\lambda_{l}-\lambda_{l'})})}
         {\displaystyle
          \sum_{l=1}^{N+1}
            \frac{\lambda_{l}^{\,L}}
                 {\prod_{\substack{1\le l'\le N+1\\l'\neq l}}\qty(\lambda_{l}-\lambda_{l'})}}.
\end{align}
Here we used~\eqref{eq:Bmatpower} to derive the final line.

\end{widetext}

\subsection{Many-body particle distribution and correlations in the two-species case}\label{subsec:many-body}
In this subsection, we consider the case of two particle species ($M=2$).  
We use Eqs.~\eqref{eq:numdens} and \eqref{eq:numcorr} derived in Sec.~\ref{subsec:distribution-correlation}.  
We calculate the steady particle number expectation $\kitaiti{\hat{n}_{j}}$, the variance  
$
  \kitaiti{\hat{n}_{j}^{2}}-\kitaiti{\hat{n}_{j}}^{2}
  = \kitaiti{\hat{n}_{j}}\qty(1 - \kitaiti{\hat{n}_{j}})
  \quad\qty(\hat{n}_{j}^{2} = \hat{n}_{j}),
$
the correlation function $\kitaiti{\hat{n}_{i}\hat{n}_{j}}$, and the covariance  
$
  \kitaiti{\hat{n}_{i}\hat{n}_{j}}
  - \kitaiti{\hat{n}_{i}}\,\kitaiti{\hat{n}_{j}}.
$
We discuss how this behavior differs from that of the standard ASEP.

\subsubsection{$\kitaiti{\hat{n}_{j}}$ and its variance in the two species case}\label{subsubsec:numdistribution}
\begin{figure}[H]
  \includegraphics[width=\columnwidth,clip]{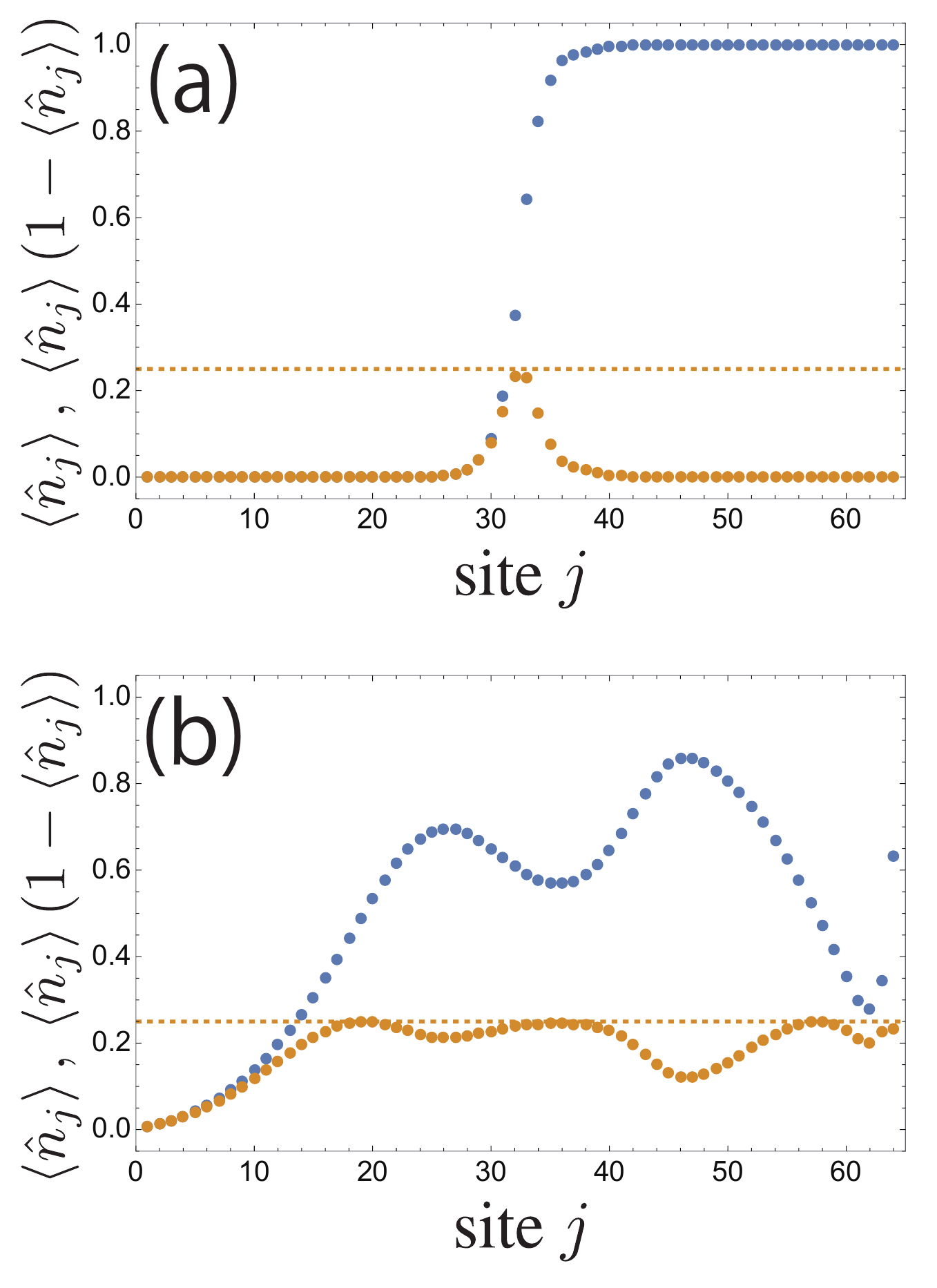}
  \caption{Particle number expectation $\kitaiti{\hat{n}_{j}}$ (blue) and variance $\kitaiti{\hat{n}^{2}_{j}}-\kitaiti{\hat{n}_{j}}^{2}$ ($\hat{n}^{2}_{j}=\hat{n}_{j}$) (orange) for the two-species case ($M=2$), with system size $L=64$ and total particle number $N=32$. Blue and orange circles denote, respectively, the particle number expectation and variance obtained by the exact MPS method. Small black circles with error bars denote kinetic Monte Carlo results (error bars are standard errors). The particle species sequence is fixed as $\chuukakko{\shortstack{1,1,1,1,2,2,2,1,2,2,2,1,2,2,1,1,1,2,1,2,1,2,2,1,2,2,2,1,2,2,2,1}}$.  
  (a): Both species hop preferentially to the right $\qty(q_{1}=1.5>1,q_{2}=1.25>1)$, producing a simple domain-wall structure typical of standard ASEP.  
  (b): Species 1 hops rightward $\qty(q_{1}=1.5>1)$ and species 2 hops leftward $\qty(q_{2}=0.75<1)$, creating a more complex and strongly correlated density profile absent in standard ASEP.
  \label{Fig-NumDensity}}
\end{figure}

Fig.~\ref{Fig-NumDensity} shows the particle number expectation and variance for a steady state $\ket{P_{\text{s.s.}}}$ under the settings given in the caption.    
Note that since $0 \le \kitaiti{\hat{n}_{j}} \le 1$, the variance ranges from $0$ to $\tfrac{1}{4}$.

Fig.~\ref{Fig-NumDensity} (a) shows the case where both species hop preferentially to the right.  
Here, the particle number distribution forms a right‑biased domain wall, just as in the standard ASEP.  
The variance is very small except at the domain‑wall interface, indicating that the steady state has little nonlocal correlation.  
This can also be inferred from the MPS expansion in Eq.~\eqref{eq:Pss}: since both $q_{1}$ and $q_{2}$ exceed 1, configurations biased to the right have exponentially larger weights, making the state close to a product state.  
Such behavior also appears in the RBC ASEP or the kink boundary condition XXZ model~\cite{koma1997spectral}, so it does not differ significantly from the standard ASEP.

Fig.~\ref{Fig-NumDensity} (b) shows the case where species $\sigma=1$ hops preferentially to the right and species $\sigma=2$ hops preferentially to the left.  
In this case, the particle number distribution exhibits a more complex landscape, markedly different from the standard ASEP.  
In particular, wide regions do not saturate at $\kitaiti{\hat{n}_{j}}=0$ or 1, leading to nonzero variance.  
This indicates that the steady state is far from a product state and has nonlocal correlations.  
Indeed, large regions have variance close to the maximum $\tfrac{1}{4}$.  
This too follows from ~\eqref{eq:Pss}, where $q_{1}(>1)$ and $q_{2}(<1)$ partially cancel among the coefficients of $\ket{P_{\text{s.s.}}}$ obtained by expanding in the real‐space basis, giving similar weights to many configurations and moving the steady state away from a product form.

\subsubsection{$\kitaiti{\hat{n}_{i}\hat{n}_{j}}$ and its cumulant in the two species case}\label{subsubsec:nunmcumlant}
\begin{figure}[H]
  \includegraphics[width=\columnwidth,clip]{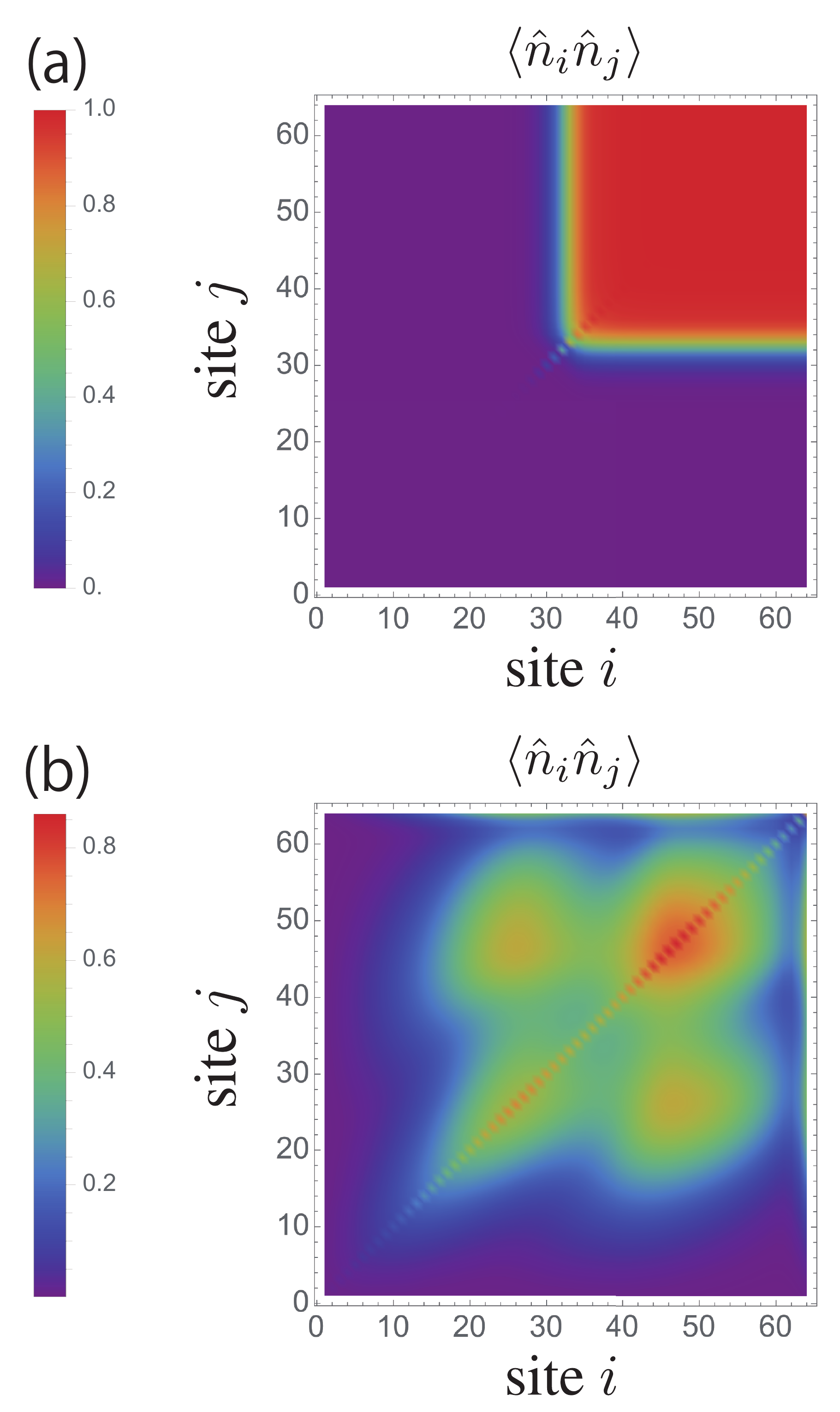}
  \caption{Color plot of the particle number correlation function $\kitaiti{\hat{n}_{i}\hat{n}_{j}}$ with axes $i$ and $j$ for $M=2$, $L=64$, $N=32$, and the configuration $\chuukakko{\shortstack{1,1,1,1,2,2,2,1,2,2,2,1,2,2,1,1,1,2,1,2,1,2,2,1,2,2,2,1,2,2,2,1}}$.  
  (a): $\qty(q_{1},q_{2})=\qty(1.5,1.25)$ case. Particles accumulate at the edges, forming a domain wall as in Fig.~\ref{Fig-NumDensity} and the ASEP steady state.  
  (b): $\qty(q_{1},q_{2})=\qty(1.5,0.75)$ case. No clear domain wall appears, showing a behavior distinct from the standard ASEP's steady state.
  \label{Fig-NumCorrelation}}
\end{figure}

\begin{figure}[H]
  \includegraphics[width=\columnwidth,clip]{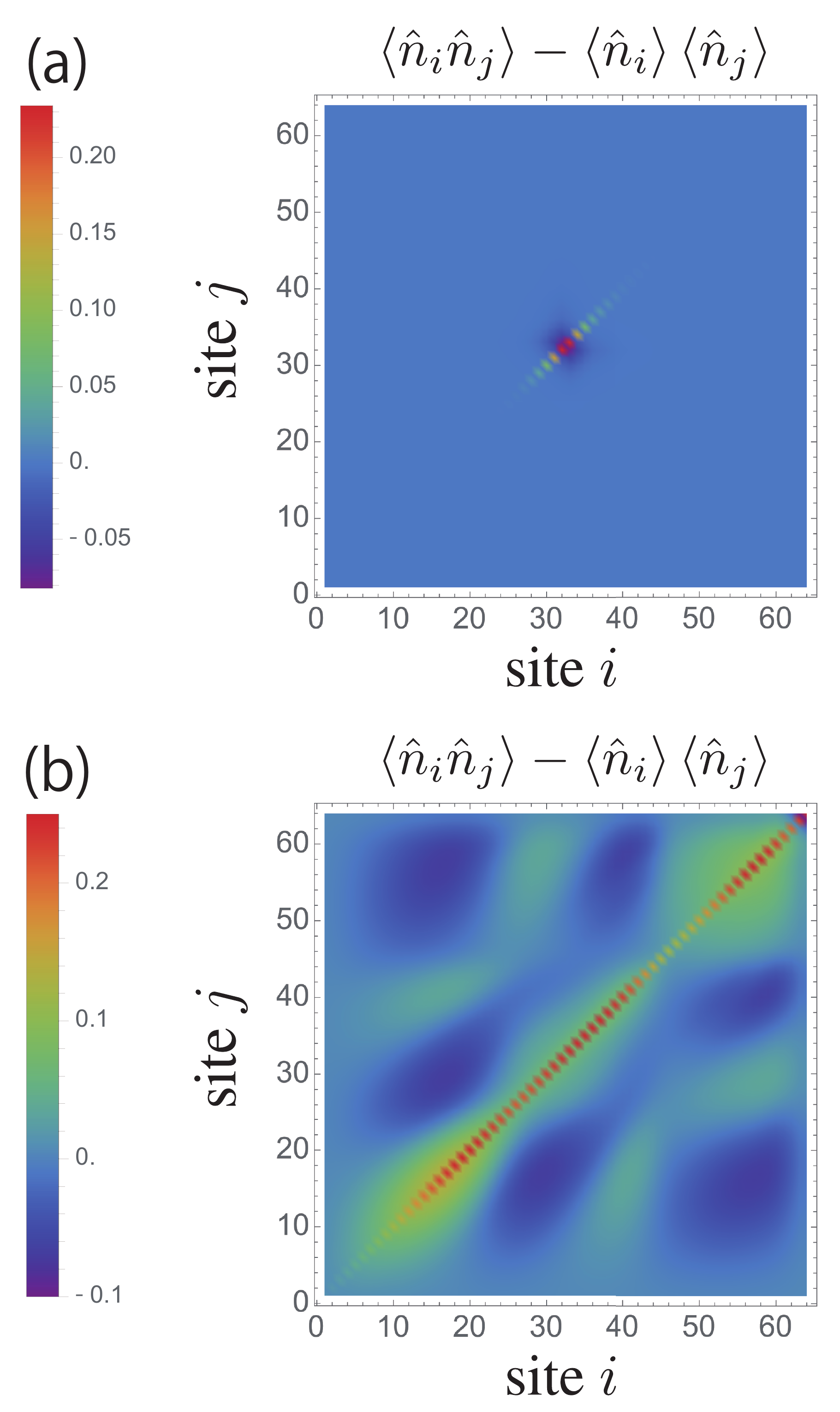}
  \caption{Color plot of the cumulant of particle number correlation $\kitaiti{\hat{n}_{i}\hat{n}_{j}}-\kitaiti{\hat{n}_{i}}\,\kitaiti{\hat{n}_{j}}$ with axes $i$ and $j$ for $M=2$, $L=64$, $N=32$, and the configuration $\chuukakko{\shortstack{1,1,1,1,2,2,2,1,2,2,2,1,2,2,1,1,1,2,1,2,1,2,2,1,2,2,2,1,2,2,2,1}}$.  
  (a): $\qty(q_{1},q_{2})=\qty(1.5,1.25)$ case. Non-zero values appear only near the center, indicating that the domain wall in Fig.~\ref{Fig-NumCorrelation} mainly arises from product densities.  
  (b): $\qty(q_{1},q_{2})=\qty(1.5,0.75)$ case. The plot shows a complex landscape with long‑range nonzero values, demonstrating nonlocal correlations in the steady state.}
  \label{Fig-NumCumulant}
\end{figure}

Fig.~\ref{Fig-NumCorrelation} and Fig.~\ref{Fig-NumCumulant} display color plots of the correlation function $\kitaiti{\hat{n}_{i}\hat{n}_{j}}$ and its cumulant $\kitaiti{\hat{n}_{i}\hat{n}_{j}}-\kitaiti{\hat{n}_{i}}\,\kitaiti{\hat{n}_{j}}$, respectively.  
The parameters and configuration are the same as in SubsubSec.~\ref{subsubsec:numdistribution} and Fig.~\ref{Fig-NumDensity}.

In (a) of both figures, both species hop preferentially to the right.  
Fig.~\ref{Fig-NumCorrelation} shows a clear domain wall structure.  
In Fig.~\ref{Fig-NumCumulant}, the cumulant is nearly zero except near the center.  
This again indicates that the steady state has little nonlocal correlation and resembles a product state.

In (b), species $\sigma=1$ hops rightward and $\sigma=2$ leftward.  
Both figures show a complex landscape very different from (a).  
Notably, Fig.~\ref{Fig-NumCumulant}(b) shows nonzero cumulant even for distant sites $i,j$, indicating system-spanning correlations.  
This confirms that the steady state has strong nonlocal correlations and is far from a product state.

\section{Gap Structure and Relaxation Times for Two‑Species System}\label{sec:gap}

In this section, we study how Hilbert-space fragmentation affects relaxation in the two-species case ($M=2$) through sector-resolved spectral gaps.  
We do not attempt to solve the full many-body spectrum in every sector.  
Instead, we combine (i) exact results in analytically tractable sectors (Sec.~\ref{subsec:XXZsectorgap}), (ii) a finite-size Knabe criterion for frustration-free chains, and (iii) exact diagonalization of the sector that is numerically observed to have the smallest nonzero gap (Sec.~\ref{subsec:minimum-gap}).  
As a relaxation-scale indicator for the Markov process generated by $W$, we define in each particle species sequence sector $S$ of the finite RBC system the gap as
\begin{align}
    \Delta\qty(S) \coloneqq E_{1}\qty(S) - E_{0}\qty(S),
\end{align}
where $E_{0}\qty(S)=0$ is the ground‑state energy and $E_{1}\qty(S)$ is the first excited‑state energy.  
This definition includes possible edge modes that would vanish under periodic boundary conditions.  
Following Ref.~\cite{levin2017markov}, in many-body Markov processes $\tau\qty(S)^{-1}$ need not exactly coincide with $\Delta\qty(S)$ because overlap factors can depend on system size.  
We therefore use $\Delta\qty(S)$ as an indicator of relaxation scales.
Even with this limitation, the sector-dependent gaps show that for the same parameters~$\qty(p_{1},p_{2})$, different initial states can lead either to rapid relaxation or to long-lived metastable behavior.

\subsection{Smallest Gap Sector}\label{subsec:minimum-gap}

From exact diagonalization over sectors at each finite size studied, we find that the smallest observed gap occurs in the one‑hole sector $N=L-1$ with particle species sequence
\begin{align}
  S_{\mathrm{min}}=\qty(\underbrace{1,\cdots,1}_{N_{1}},\underbrace{2,\cdots,2}_{N_{2}}),
  \quad N_{1}+N_{2}=L-1,
\end{align}
where $N_{1}$ and $N_{2}$ are chosen to minimize the gap.  
We denote this minimal gap by $\Delta_{\mathrm{min}}$.  
The restriction of $H$ to this sector is an $L\times L$ matrix given in Appendix~\ref{app:Hamrep}.  
Fig.~\ref{Fig-mingapsector} shows how $\Delta_{\mathrm{min}}$ scales with $L$ for parameters $\qty(p_{1},p_{2})$ (see Appendix~\ref{app:expsmallnumerical} for numerical data).

In the region $p_{1}<\hf,p_{2}<\hf$ or $p_{1}>\hf,p_{2}>\hf$ (red in Fig.~\ref{Fig-mingapsector}), sampled values of $\Delta_{\mathrm{min}}$ at $L=1024$ exceed the threshold function $G\qty(L)$ defined in Ref.~\cite{lemm2019spectral}:
\begin{align}
\label{eq:threshold}
    &G\qty(L)\coloneqq\frac{1 + a_{L}^{2} b_{L}}{L - 1 + L^{3/2} a_{L}},\nt
    &a_{L}\coloneqq -\frac{L-1}{L^{3/2}}
       + \sqrt{\qty(\frac{L-1}{L^{3/2}})^2
           + \frac{1}{b_{L}}},\nt
    &b_{L}\coloneqq \frac{6L^{3}}{\qty(L-1)\qty(L-2)\qty(L-3)}.
\end{align}
Knabe's criterion compares the finite-size gap with the threshold function $G(L)$.  
If $\Delta_{\mathrm{min}}(L)>G(L)$ at a parameter point, then the corresponding infinite system is gapped.
Since $G(L=1024)=0.000138884$ and $\Delta_{\mathrm{min}}$ in all sample points satisfy $\Delta_{\mathrm{min}}>G(1024)$, the sampled points are rigorously gapped and strongly support an $O(1)$ gap throughout the red region (see Appendix~\ref{app:gappedproof}).  
This agrees with the gapped Ising‑like ferro XXZ line on $p_{1}=p_{2}$ (green) contained in the red region.

On the lines $p_{1}=\hf$ or $p_{2}=\hf$ (cyan), $\Delta_{\mathrm{min}}$ closes as a power law, $\Delta_{\mathrm{min}}\sim L^{-z}$ with $z\ge2$ (see Appendix~\ref{app:expsmallnumerical}).  
This matches the bound $z\ge2$ for frustration‑free gapless models~\cite{PhysRevB.110.195140,masaoka2024rigorouslowerbounddynamic} and the Nambu–Goldstone magnon at $p_{1}=p_{2}=\hf$ (Heisenberg limit).

In the region $p_{1}<\hf<p_{2}$ or $p_{2}<\hf<p_{1}$ (purple), $\Delta_{\mathrm{min}}$ is numerically observed to close exponentially with $L$ (see Appendix~\ref{app:expsmallnumerical}) like Motzkin chain~\cite{Levine_2017}, Fredkin chain~\cite{Udagawa_2017,Zhang_2017} and related models.  
This behavior is consistent with gap closing in the thermodynamic limit, but the infinite-volume spectral structure (including possible steady-state degeneracy) remains open.
For large but finite $L$, the observed scaling indicates exponentially long relaxation times and metastable behavior.  
This phenomenon is absent in the ordinary RBC ASEP and is a unique feature of the no-passing ASEP in this parameter region.

\begin{figure}[H]
  \includegraphics[width=\columnwidth,clip]{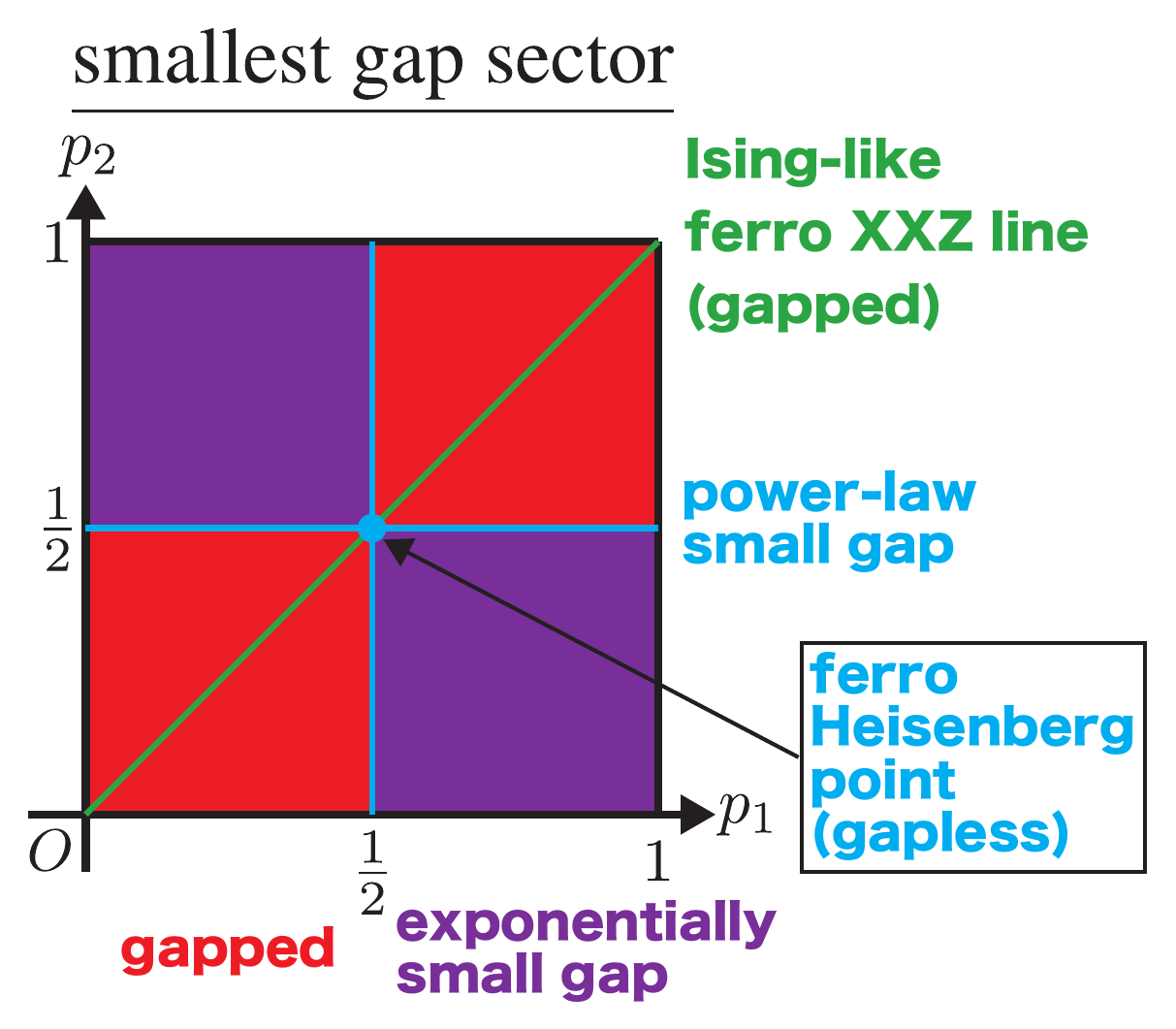}
  \caption{Scaling of the minimal gap $\Delta_{\mathrm{min}}$ with system size $L$ in the sector with the minimal gap.  
  In the red region ($p_{1}<\hf$, $p_{2}<\hf$ or $p_{1}>\hf$, $p_{2}>\hf$), sampled points satisfy the Knabe threshold~\cite{lemm2019spectral}, consistent with a finite-gap region.  
  This region contains the Ising‑like ferro XXZ line $p_{1}=p_{2}$ (green).  
  On the cyan lines ($p_{1}=\hf$ or $p_{2}=\hf$), the gap closes as a power law with $L$.  
  In the purple region ($p_{1}<\hf<p_{2}$ or $p_{2}<\hf<p_{1}$), the gap is numerically observed to close exponentially with $L$.  
  The central point $p_{1}=p_{2}=\hf$ is the gapless ferro Heisenberg limit.  
  \label{Fig-mingapsector}}
\end{figure}

\subsection{Single‑Species XXZ Sector Gaps}\label{subsec:XXZsectorgap}

\begin{figure}[H]
  \includegraphics[width=\columnwidth,clip]{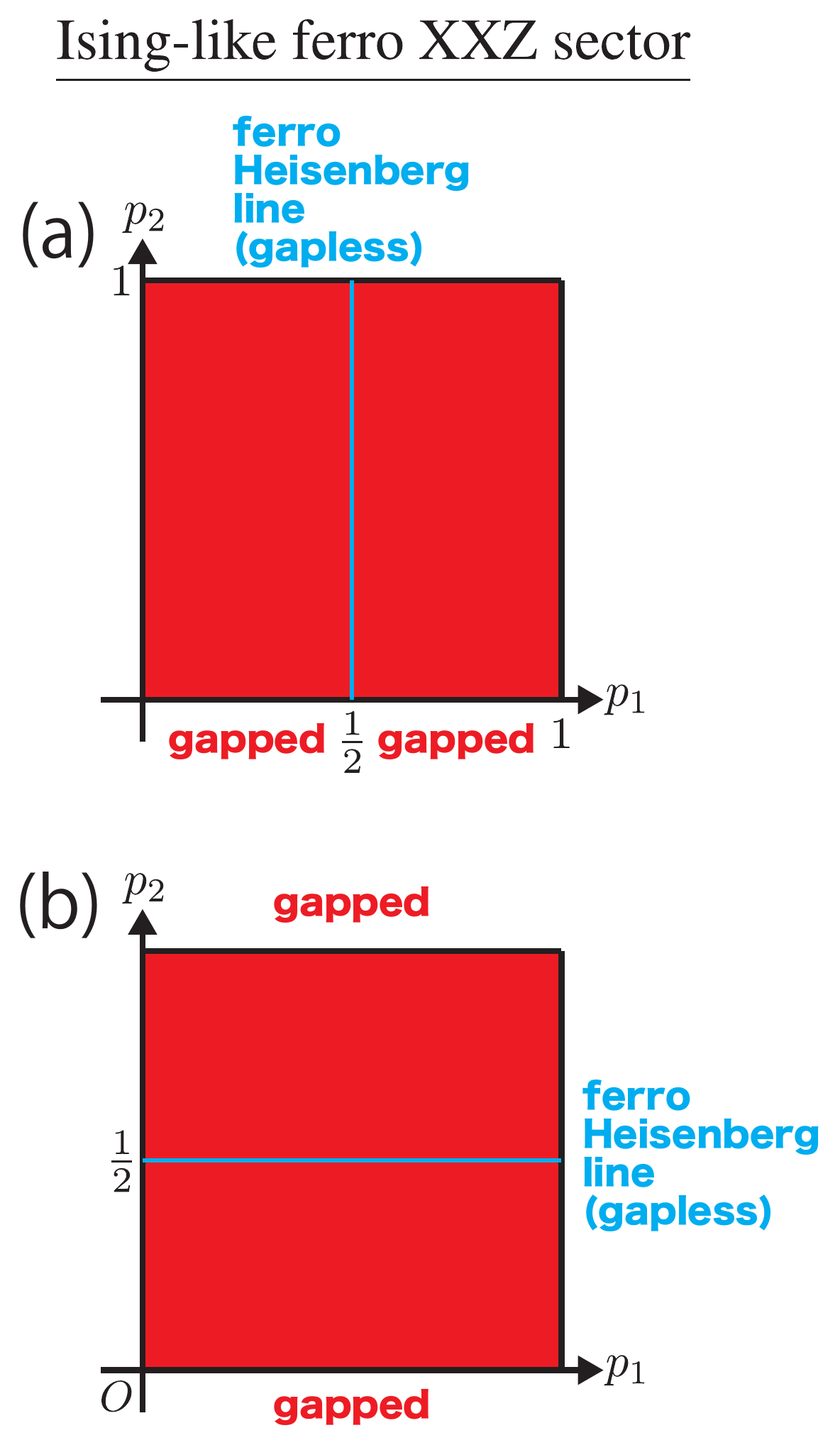}
  \caption{Gaps in single‑species sectors from \eqref{eq:normalASEP} or \eqref{eq:XXZmap}.  
  (a): Only species $\sigma=1$.  The gap is finite for $p_{1}\neq\hf$ and closes only at $p_{1}=\hf$.  
  (b): Only species $\sigma=2$.  The gap is finite for $p_{2}\neq\hf$ and closes only at $p_{2}=\hf$.  
  \label{Fig-XXZsector}}
\end{figure}

Sectors with only one species $\sigma$ reduce to the Ising‑like ferro XXZ model (see Appendix.~\ref{app:sector}).  
These sectors further decompose by total particle number $\hat{N}$.  
According to Ref.~\cite{koma1997spectral}, under kink‑boundary conditions, each sub‑sector’s gap is bounded below by the single‑particle gap
\begin{align}
    \Delta\qty(\chuukakko{\sigma})
    &=1-2\sqrt{p_{\sigma}\qty(1-p_{\sigma})}\cos\qty(\tfrac{\pi}{L})\nt
    &\xrightarrow{L\to\infty}
    1-2\sqrt{p_{\sigma}\qty(1-p_{\sigma})}.
\end{align}
Thus for $p_{\sigma}\neq\hf$, all single‑species sectors are gapped (red in Fig.~\ref{Fig-XXZsector}).
This is consistent with the result of SubSec.~\ref{subsec:minimum-gap}.
At $p_{\sigma}=\hf$ (cyan), the Heisenberg limit is gapless.  
Fig.~\ref{Fig-XXZsector} illustrates this behavior for $\sigma=1$ and $\sigma=2$.

\subsection{Global Gap Structure and Initial‑State Dependence}\label{subsec:expgap}

Combining the results of Subsecs.~\ref{subsec:minimum-gap} and \ref{subsec:XXZsectorgap}, we summarize the present gap landscape and its implications for initial-state dependence.  
Single-species XXZ sectors are treated analytically, while the red-cyan-purple map is inferred from exact diagonalization of the numerically minimal-gap sector together with Knabe-threshold checks.
 
Sectors with exponentially small gaps can trap the system in metastable states for exponentially long times, significantly deviating from typical ASEP behavior; sectors with finite gaps relax rapidly.

In the red region of Fig.~\ref{Fig-mingapsector}, 
sampled values of $\Delta_{\mathrm{min}}$ satisfy the Knabe criterion at $L=1024$, indicating an $O(1)$ bulk gap at those points and suggesting weak initial-state dependence there.  
In the purple region, the observed exponential closing of $\Delta_{\mathrm{min}}$ indicates that at least some sectors relax extremely slowly.  
Although single-species XXZ sectors remain gapped there, a full sector-by-sector classification is still open.  
Accordingly, when the two species tend to hop in opposite directions, relaxation can depend strongly on the initial ordering, leading either to fast relaxation or to metastable dynamics.

Since the purple region of Fig.~\ref{Fig-mingapsector} also shows strong long-range steady-state correlations (see SubSec.~\ref{subsec:many-body}), determining gaps in each sector is an important open problem.  
In this frustration‑free one‑dimensional system, spontaneous breaking of $\mathrm{U}\qty(1)$ symmetry and emergent Nambu–Goldstone modes may also cause gapless behavior ~\cite{PhysRevLett.133.176001}.  
Variational bounds with min-max principle, \emph{nonrelativistic} field‑theory approaches ~\cite{PhysRevB.110.195140,masaoka2024rigorouslowerbounddynamic}, and the study of bulk gaps that exclude edge modes are promising directions for future work.

\section{Conclusion}\label{sec:conclusion}
In this work, we introduced a multispecies no-passing ASEP on a one-dimensional reflecting (closed) chain and clarified how congestion structures emerge from competing hopping asymmetries.
Combining exact analysis and numerics, we showed that Hilbert-space fragmentation produces strongly initial-order-dependent relaxation and can generate metastability, as detailed in Secs.~\ref{sec:gap} and \ref{subsec:many-body}.

To state clearly what is established here for the first time in this setting: (i) we identify sector-dependent relaxation, with distinct gap scaling across fixed particle species sequence sectors; (ii) we derive closed-form one- and two-point steady correlations from the exact MPS-transfer-matrix framework; and (iii) we identify a metastable regime in the two-species model associated with exponentially small sector gaps.

Focusing on the two-species case, we discovered that when both species hop preferentially in the same direction the system exhibits a familiar domain-wall steady profile with only weak correlations.
This behavior is much like the standard ASEP.

In contrast, in the regions in which the particles' preferred hopping directions alternate, the steady state develops a richly structured density landscape with strong long-range correlations (Sec.~\ref{subsec:many-body}, Figs.~\ref{Fig-NumDensity}, \ref{Fig-NumCorrelation}, \ref{Fig-NumCumulant}).
Our spectral-gap analysis (Sec.~\ref{sec:gap}, Fig.~\ref{Fig-mingapsector}) gives exact results in special sectors and numerical evidence that some particle species sequence sectors remain gapped while others show exponentially small gaps over the studied sizes.
This sector-dependent gap structure is consistent with metastable relaxation dynamics absent in the usual exclusion processes.

Future work includes a rigorous characterization of bulk-gap behavior in the thermodynamic limit and of sector degeneracy as $L\to\infty$.
It is also important to clarify fluctuation universality, including possible Kardar-Parisi-Zhang scaling, and to investigate thermalization properties of the Hermitianized Hamiltonian $H$ under Hilbert-space fragmentation constraints.

\section*{Acknowledgments}
\label{ACKNOWLEDGEMENT}
We thank H. Hayakawa, H. Katsura, M. Kunimi, T. Saito, T. Sasamoto, T. Takama, and K. Totsuka for their helpful discussions.
This model was inspired by the display at AOAO Sapporo Aquarium in Sapporo, Japan~\cite{aoao_sapporo}, which emphasized the mating behavior of Razorfish.
We are especially grateful to AOAO Sapporo Aquarium for showcasing Razorfish in such an innovative manner.
U.M. was supported by the Japan Science and Technology Agency (JST) SPRING, Grant No. JPMJSP2110, and the Japan Science and Technology Agency (JST) CREST, Grant No. JPMJCR19T2.

\section*{Data Availability}

The data and code that support the findings of this article are openly available in the GitHub repository no-passing-ASEP~\cite{no_passing_asep_code}.

\appendix
\onecolumngrid
\section{Summary of Symbols}\label{app:symbols}

For quick reference, Table~\ref{tab:symbol-summary} lists the symbols in the order they first appear in the paper, together with their defining equations.

\begin{center}
\footnotesize
\setlength{\tabcolsep}{3pt}
\renewcommand{\arraystretch}{0.93}
\tablecaption{Summary of symbols and definitions used throughout the paper. Symbols are listed in the order in which they first appear.\label{tab:symbol-summary}}
\tablehead{\toprule Symbol & Definition or meaning (in order of appearance)\\ \midrule}
\tabletail{\bottomrule}
\begin{supertabular}{p{0.18\textwidth}p{0.76\textwidth}}
$L$ & One-dimensional chain length (number of sites).\\
$M$ & Number of particle species.\\
$N$ & Particle number in a configuration, $0\le N\le L$.\\
$(j_{1},\ldots,j_{N})$ & Ordered particle positions with $1\le j_{1}<\cdots<j_{N}\le L$.\\
$(\sigma_{1},\ldots,\sigma_{N})$ & Ordered species labels with $\sigma_{n}\in\{1,\ldots,M\}$.\\
$C$ & Configuration label: $C:=((j_{1},\sigma_{1}),\ldots,(j_{N},\sigma_{N}))$.\\
$\mathcal{H}_{j}$ & Local Hilbert space: $\mathcal{H}_{j}=\mathrm{span}(\ket{0}_{j},\ket{1}_{j},\ldots,\ket{M}_{j})$.\\
$\ket{0}_{j},\ket{\sigma}_{j}$ & Local basis states (hole or one particle of species $\sigma$) at site $j$.\\
$b_{j,\sigma},\hc{b}_{j,\sigma}$ & Hard-core ladder operators: $b_{j,\sigma}:=\ket{0}\bra{\sigma}_{j}$, $\hc{b}_{j,\sigma}:=\ket{\sigma}\bra{0}_{j}$.\\
$\hat{n}_{j,\sigma}$ & Species-resolved density: $\hat{n}_{j,\sigma}:=\ket{\sigma}\bra{\sigma}_{j}=\hc{b}_{j,\sigma}b_{j,\sigma}$.\\
$\hat{n}_{j}$ & Total local density: $\hat{n}_{j}:=\sum_{\sigma=1}^{M}\hat{n}_{j,\sigma}=1-\ket{0}\bra{0}_{j}$.\\
$\mathcal{H}$ & Full Hilbert space: $\mathcal{H}:=\bigotimes_{j=1}^{L}\mathcal{H}_{j}$.\\
$\ket{\Omega}$ & Vacuum state: $\ket{\Omega}:=\bigotimes_{j=1}^{L}\ket{0}_{j}=\ket{0,\ldots,0}$.\\
$\ket{C}$ & Basis vector for configuration $C$: $\ket{C}:=\hc{b}_{j_{1},\sigma_{1}}\cdots\hc{b}_{j_{N},\sigma_{N}}\ket{\Omega}$.\\
$\pi(t;C)$ & Probability of configuration $C$ at time $t$.\\
$W_{C'\leftarrow C}$ & Markov matrix element with sign convention: transition rate is $-W_{C'\leftarrow C}$ ($C'\neq C$).\\
$W_{C\leftarrow C}$ & Diagonal term: $-W_{C\leftarrow C}:=\sum_{C'\neq C}W_{C'\leftarrow C}$, and $\sum_{C'}W_{C'\leftarrow C}=0$.\\
$\ket{P(t)}$ & Probability vector: $\ket{P(t)}:=\sum_{C}\pi(t;C)\ket{C}$.\\
$W$ & Markov generator: $W:=\sum_{C,C'}W_{C'\leftarrow C}\ket{C'}\bra{C}$.\\
$W_{j,j+1}$ & Local bond generator, with $W=\sum_{j=1}^{L-1}W_{j,j+1}$ [Eq.~\eqref{eq:nonHHam}].\\
$p_{\sigma}$ & Hop parameter of species $\sigma$: left hop $\ket{0,\sigma}\to\ket{\sigma,0}$ at rate $p_{\sigma}$, right hop $\ket{\sigma,0}\to\ket{0,\sigma}$ at rate $1-p_{\sigma}$.\\
$q_{\sigma}$ & Bias parameter: $q_{\sigma}:=\sqrt{(1-p_{\sigma})/p_{\sigma}}$.\\
$S_{\sigma},S,S^{-1}$ & Similarity operators: $S:=\prod_{\sigma=1}^{M}S_{\sigma}$, $S_{\sigma}:=q_{\sigma}^{-\sum_{j=1}^{L}j\hat{n}_{j,\sigma}}$, $S^{-1}=\prod_{j=1}^{L}\left[(1-\hat{n}_{j})+\sum_{\sigma=1}^{M}q_{\sigma}^{j}\hat{n}_{j,\sigma}\right]$.\\
$H,h_{j,j+1}$ & Hermitianized Hamiltonian: $H:=SWS^{-1}=\sum_{j=1}^{L-1}h_{j,j+1}$ [Eq.~\eqref{eq:HermHam}].\\
$\hat{N}_{\sigma},\hat{N}$ & Conserved numbers: $\hat{N}_{\sigma}:=\sum_{j=1}^{L}\hat{n}_{j,\sigma}$, $\hat{N}:=\sum_{j=1}^{L}\hat{n}_{j}$.\\
$P$ & Parity unitary: $\hc{P}b_{j,\sigma}P=b_{L+1-j,\sigma}$.\\
$U_{\sigma_{1},\sigma_{2}}$ & Species-swap unitary [Eq.~\eqref{eq:12swap}] exchanging labels $\sigma_{1},\sigma_{2}$ in $H$.\\
$S$ & Conserved particle species sequence: $S=(\sigma_{1},\ldots,\sigma_{N})$.\\
$\mathcal{M}$ & Sequence operator: $\mathcal{M}:=\sum_{j=1}^{L}M^{\hat{n}_{1}+\cdots+\hat{n}_{j-1}}\sum_{\sigma=1}^{M}\sigma\hat{n}_{j,\sigma}$ [Eq.~\eqref{eq:sequenceop}].\\
$H^{S},H^{\mathcal{M}=m}$ & Block Hamiltonians in a fixed sequence sector or fixed eigenvalue sector.\\
$\pi_{\mathrm{s.s.}}(C)$ & Sector steady distribution: $\pi_{\mathrm{s.s.}}(C)\propto\prod_{n=1}^{N}q_{\sigma_{n}}^{2j_{n}}$.\\
$\ket{P_{\mathrm{s.s.}}}$ & Normalized steady state [Eq.~\eqref{eq:Pss}].\\
$\ket{Q_{-}(\sigma)}_{j,j+1}$ & Two-site projector state: $\sqrt{1-p_{\sigma}}\ket{\sigma,0}_{j,j+1}-\sqrt{p_{\sigma}}\ket{0,\sigma}_{j,j+1}$.\\
$\ket{Q_{+}(\sigma)}_{j,j+1}$ & Two-site zero-mode state: $\sqrt{p_{\sigma}}\ket{\sigma,0}_{j,j+1}+\sqrt{1-p_{\sigma}}\ket{0,\sigma}_{j,j+1}$.\\
$\lambda_{m}$ & MPS weight [Eq.~\eqref{eq:lambdas}]: $\lambda_{m}=(q_{\sigma_{m}}\cdots q_{\sigma_{N}})^{2}$ ($m\le N$), $\lambda_{N+1}=1$.\\
$\mathsf{A}_{j}$ & Ground-state MPS tensor: $(\mathsf{A}_{j})_{m,m}=\sqrt{\lambda_{m}}\ket{0}_{j}$, $(\mathsf{A}_{j})_{m,m+1}=\ket{\sigma_{m}}_{j}$.\\
$\ket{\bar{Q}_{+}(\sigma)}_{j,j+1}$ & Kernel vector: $\ket{\bar{Q}_{+}(\sigma)}_{j,j+1}:=\ket{\sigma,0}_{j,j+1}+q_{\sigma}\ket{0,\sigma}_{j,j+1}$.\\
$\bmm{e}_{m}$ & Standard basis vector in the $(N+1)$-dimensional auxiliary space.\\
$\ket{\Psi'_{\mathrm{g.s.}}}$ & Unnormalized ground state: $\ket{\Psi'_{\mathrm{g.s.}}}=\bmm{e}_{1}^{\top}\mathsf{A}_{1}\cdots\mathsf{A}_{L}\bmm{e}_{N+1}$ [Eq.~\eqref{eq:Psidash}].\\
$\ket{P'_{\mathrm{s.s.}}}$ & Unnormalized steady state: $\ket{P'_{\mathrm{s.s.}}}:=S^{-1}\ket{\Psi'_{\mathrm{g.s.}}}=\bmm{e}_{1}^{\top}\mathsf{B}_{1}\cdots\mathsf{B}_{L}\bmm{e}_{N+1}$.\\
$\mathsf{B}_{j}$ & Steady-state MPS tensor: $(\mathsf{B}_{j})_{m,m}=\lambda_{m}\ket{0}_{j}$, $(\mathsf{B}_{j})_{m,m+1}=\ket{\sigma_{m}}_{j}$.\\
$\mathsf{M}_{j}$ & Summation bra used in transfer reduction: $\mathsf{M}_{j}:=\bra{0}_{j}+\sum_{\sigma=1}^{M}\bra{\sigma}_{j}$.\\
$\Tilde{\mathsf{B}}$ & Transfer matrix: $\Tilde{\mathsf{B}}:=\bra{0}_{j}\mathsf{B}_{j}+\sum_{\sigma=1}^{M}\bra{\sigma}_{j}\mathsf{B}_{j}$ [Eq.~\eqref{eq:TransferB}].\\
$\mathcal{N}$ & Normalization constant: $\mathcal{N}:=\sum_{\{C\}}\braket{\{C\}}{P'_{\mathrm{s.s.}}}=\bmm{e}_{1}^{\top}\Tilde{\mathsf{B}}^{L}\bmm{e}_{N+1}$.\\
$\Tilde{\mathsf{N}}$ & Density insertion transfer matrix: $\Tilde{\mathsf{N}}:=\bra{0}\hat{n}\mathsf{B}+\sum_{\sigma=1}^{M}\bra{\sigma}\hat{n}\mathsf{B}=\sum_{m=1}^{N}\bmm{e}_{m}\bmm{e}_{m+1}^{\top}$ [Eq.~\eqref{eq:TransferN}].\\
$\kitaiti{A}$ & Steady-state expectation value: $\kitaiti{A}=\sum_{\{C\}}\bra{\{C\}}A\ket{P_{\mathrm{s.s.}}}$.\\
$\kitaiti{\hat{n}_{j}}$ & One-point density [Eq.~\eqref{eq:numdens}]: $\kitaiti{\hat{n}_{j}}=\mathcal{N}^{-1}\bmm{e}_{1}^{\top}\Tilde{\mathsf{B}}^{j-1}\Tilde{\mathsf{N}}\Tilde{\mathsf{B}}^{L-j}\bmm{e}_{N+1}$.\\
$\kitaiti{\hat{n}_{i}\hat{n}_{j}}$ & Two-point function via two $\Tilde{\mathsf{N}}$ insertions in the transfer product.\\
$\Delta(S)$ & Sector gap: $\Delta(S):=E_{1}(S)-E_{0}(S)$ with $E_{0}=0$.\\
$\tau(S)$ & Relaxation-time indicator per sector (used with caveat that $\tau^{-1}$ need not exactly equal $\Delta$).\\
$\Delta_{\mathrm{min}}$ & Smallest sector gap at fixed parameters and size: $\Delta_{\mathrm{min}}:=\min_{S}\Delta(S)$.\\
$G(L),a_{L},b_{L}$ & Knabe's threshold functions: $G(L):=\frac{1+a_{L}^{2}b_{L}}{L-1+L^{3/2}a_{L}}$, $a_{L}:=-\frac{L-1}{L^{3/2}}+\sqrt{\qty(\frac{L-1}{L^{3/2}})^{2}+\frac{1}{b_{L}}}$, $b_{L}:=\frac{6L^{3}}{(L-1)(L-2)(L-3)}$.\\
$W^{\tau},W_{j,j+1}^{\tau}$ & Single-species sector generator and local term [Eq.~\eqref{eq:normalASEP}].\\
$H^{\tau},h_{j,j+1}^{\tau}$ & Hermitianized single-species Hamiltonian and local term: $H^{\tau}:=SW^{\tau}S^{-1}=\sum_{j=1}^{L-1}h_{j,j+1}^{\tau}$.\\
$\sigma_{j}^{\pm}$ & Spin operators used in XXZ mapping: $\sigma_{j}^{\pm}:=\frac{1}{2}(\sigma_{j}^{x}\pm i\sigma_{j}^{y})$.\\
$\hc{B}_{\sigma}$ & Algebraic ground-state generator: $\hc{B}_{\sigma}:=\sum_{j=1}^{L}q_{\sigma}^{j}\hc{b}_{j,\sigma}$ [Eq.~\eqref{eq:Bgenerator}].\\
$\ket{(r_{1},\ldots,r_{M})}$ & Algebraic state: $\ket{(r_{1},\ldots,r_{M})}:=(\hc{B}_{1})^{r_{1}}\cdots(\hc{B}_{M})^{r_{M}}\ket{\Omega}$ with $N=\sum_{\sigma=1}^{M}r_{\sigma}$.\\
$\hat{P}(S)$ & Projector onto fixed sequence sector $S$ (explicit MPO form in Appendix~\ref{app:algebraic}).\\
$t_{1},t_{2}$ & Smallest-gap-sector hopping scales: $t_{1}:=\sqrt{p_{1}(1-p_{1})}$, $t_{2}:=\sqrt{p_{2}(1-p_{2})}$.\\
\end{supertabular}
\normalsize
\end{center}
\twocolumngrid

\section{Sector Reducing to the Standard ASEP}\label{app:sector}
Consider the sector containing particles of only a single species $\tau$.  
In this case, no interactions occur between different species, so the system reduces to the standard ASEP.  
After the similarity transformation, the Hermitian Hamiltonian \eqref{eq:HermHam} also reduces to the ferromagnetic XXZ model of the standard ASEP.  
Since $b_{j,\sigma}=0\ \qty(\sigma\neq\tau)$ and $\hat{n}_{j}=\hat{n}_{j,\tau}$ in this sector, \eqref{eq:nonHHam} and \eqref{eq:HermHam} for $W$ and $H$ reduce to:
\begin{align}
\label{eq:normalASEP}
    &W^{\tau}=\sum_{j=1}^{L-1} W_{j,j+1}^{\tau}, \nt
    &W_{j,j+1}^{\tau}=\nt
    -& \qty[p_{\tau}\hc{b}_{j,\tau}b_{j+1,\tau} +\qty(1-p_{\tau})\hc{b}_{j+1,\tau}b_{j,\tau}]\nt
    +&p_{\tau}\qty(1-\hat{n}_{j,\tau})\hat{n}_{j+1,\tau}
      +\qty(1-p_{\tau})\hat{n}_{j,\tau}\qty(1-\hat{n}_{j+1,\tau}),\nt
    &H^{\tau}\coloneqq S\,W^{\tau}S^{-1}
      =\sum_{j=1}^{L-1} h_{j,j+1}^{\tau}, \nt
    &h_{j,j+1}^{\tau}=\nt
    -& \sqrt{p_{\tau}\qty(1-p_{\tau})}\qty(\hc{b}_{j,\tau}b_{j+1,\tau}
      +\hc{b}_{j+1,\tau}b_{j,\tau})\nt
    +&p_{\tau}\qty(1-\hat{n}_{j,\tau})\hat{n}_{j+1,\tau}
      +\qty(1-p_{\tau})\hat{n}_{j,\tau}\qty(1-\hat{n}_{j+1,\tau}).
\end{align}
Here, $W^{\tau}$ is the transition matrix of the standard ASEP.  
Its Hermitianized version is $H^{\tau}$.  
By identifying $b_{j,\tau}=\sigma_{j}^{-}$ and $\hc{b}_{j,\tau}=\sigma_{j}^{+}$, with 
$\sigma_{j}^{\pm}\coloneqq\hf\qty(\sigma_{j}^{x}\pm\im\sigma_{j}^{y})$ ($\sx,\sy \text{ and } \sz$ are standard Pauli matrices with eigenvalues $\pm 1$), we obtain a ferromagnetic XXZ model with kink-boundary condition ~\cite{koma1997spectral}:
\begin{align}
\label{eq:XXZmap}
    &W_{j,j+1}^{\tau}=\nt
    -& \frac{\sqrt{p_{\tau}\qty(1-p_{\tau})}}{2}\qty(\sx_{j}\sx_{j+1}+\sy_{j}\sy_{j+1})
      +\frac{1}{4}\qty(1-\sz_{j}\sz_{j+1})\nt
    -&\frac{p_{\tau}-\hf}{2}\qty(\sz_{j}-\sz_{j+1}).
\end{align}
This XXZ model is of Ising-like because $0<\sqrt{p_{\tau}\qty(1-p_{\tau})}\leq\hf$.  
At $p_{\tau}=\hf$, the symmetric simple exclusion process case, it becomes the isotropic ferromagnetic Heisenberg model.  
This mapping to the XXZ model is important for dynamics and Hilbert space fragmentation \cite{miura_future}, since the XXZ model is Bethe integrable and thus defines an integrable sector.

\section{Algebraic Construction of ground states}\label{app:algebraic}

In Sec.~\ref{subsec:ssMPS}, we constructed the MPS form of the ground states $\ket{\Psi_{\text{g.s.}}}$ by heuristic arguments.  
Here we give an algebraic construction starting from the trivial vacuum $\ket{\Omega}$ with $H\ket{\Omega}=0$, using raising operators akin to a quantum group (cf.\ Appendix G of Ref.~\cite{PhysRevB.98.155119}).

We define $M$ raising operators $\hc{B}_{\sigma}$ ($\sigma=1,\dots,M$) by Ref.~\cite{PhysRevB.98.155119}:
\begin{align}
  \label{eq:Bgenerator}
  \hc{B}_{\sigma}
  &\coloneqq \sum_{j=1}^{L} q_{\sigma}^{j}\,\hc{b}_{j,\sigma},\\
  \label{eq:BgeneratorMPO}
  &= \qty(1,0)\,
    \begin{pmatrix}
      q_{\sigma} & \hc{b}_{1,\sigma} \\
      0 & 1
    \end{pmatrix}
    \cdots
    \begin{pmatrix}
      q_{\sigma} & \hc{b}_{L,\sigma} \\
      0 & 1
    \end{pmatrix}
    \begin{pmatrix}
      0 \\ 1
    \end{pmatrix}.
\end{align}
One verifies
\begin{align}
  \label{eq:HB0}
  &H\hc{B}_{\sigma}\ket{\Omega}
  = \com{H}{\hc{B}_{\sigma}}\ket{\Omega} = 0,\\
  \label{eq:HBBcom0}
  &\com{\com{H}{\hc{B}_{\sigma_{1}}}}{\hc{B}_{\sigma_{2}}}=0.
\end{align}

Because $H$ is positive semidefinite and frustration‑free, all ground states have energy zero.  Let $\ket{\Psi_{\text{g.s.}}}$ be any ground state so that $H\ket{\Psi_{\text{g.s.}}}=0$.  Then from \eqref{eq:HBBcom0} we get
\begin{align}
  \label{eq:recursion}
  H\,\hc{B}_{\sigma_{1}}\hc{B}_{\sigma_{2}}\ket{\Psi_{\text{g.s.}}}
  &= \hc{B}_{\sigma_{1}}\,H\,\hc{B}_{\sigma_{2}}\ket{\Psi_{\text{g.s.}}}\\
  \label{eq:recursion-cont}
  &\quad + \hc{B}_{\sigma_{2}}\,H\,\hc{B}_{\sigma_{1}}\ket{\Psi_{\text{g.s.}}}.
\end{align}
Taking $\ket{\Psi_{\text{g.s.}}}=\ket{\Omega}$ in \eqref{eq:recursion} and using \eqref{eq:HB0} gives
\begin{align}
  \label{eq:recursion1}
  H\,\hc{B}_{\sigma_{1}}\hc{B}_{\sigma_{2}}\ket{\Omega}=0
  \quad\qty(\forall\,\sigma_{1},\sigma_{2}).
\end{align}
By iterating this argument, one shows
\begin{align}
  \label{eq:preGS}
  H\,\hc{B}_{\sigma_{1}}\cdots\hc{B}_{\sigma_{N}}\ket{\Omega}=0
  \quad\qty(\forall\,\sigma_{1},\dots,\sigma_{N}).
\end{align}

Hence any nonzero
\begin{align}
  \hc{B}_{\sigma_{1}}\cdots\hc{B}_{\sigma_{N}}\ket{\Omega}
\end{align}
is a ground state.  We introduce the notation
\begin{align}
  \ket{\qty(r_{1},\dots,r_{M})}
  \coloneqq \qty(\hc{B}_{1})^{r_{1}}
               \cdots
               \qty(\hc{B}_{M})^{r_{M}}
               \ket{\Omega},
\end{align}
with $N=r_{1}+\cdots+r_{M}\le L$, which is nonzero ground state of $H$ by the hard-core condition $\hc{b}_{j,\sigma_{1}}\hc{b}_{j,\sigma_{2}}=0$.

To project onto a fixed particle species sequence sector $S$ (an eigensector of $\mathcal{M}$), we use the MPO projection operators
\begin{widetext}
    \begin{align}
    &\hat{P}\qty(S)\nt
    &=\sum_{1\leq j_{1}<\cdots<j_{N}\leq L}
    \ket{0,\cdots,0,\underset{j_{1}}{\sigma_{1}},0,\cdots,0,\underset{j_{N}}{\sigma_{N}},0,\cdots,0}
    \bra{0,\cdots,0,\underset{j_{1}}{\sigma_{1}},0,\cdots,0,\underset{j_{N}}{\sigma_{N}},0,\cdots,0}\nt
    &=\bmm{e}_{1}^{\top}
    \begin{pmatrix}
\ket{0}\bra{0}       & \ket{\sigma_{1}}\bra{\sigma_{1}} &               &               &               \\
                 & \ket{0}\bra{0}   & \ket{\sigma_{2}}\bra{\sigma_{2}} &            &               \\
                 &                  & \ddots        & \ddots        &               \\
                 &                  &               & \ket{0}\bra{0}  & \ket{\sigma_{N}}\bra{\sigma_{N}}\\
                 &                  &               &               & \ket{0}\bra{0}
    \end{pmatrix}_{1}
    \cdots
    \begin{pmatrix}
\ket{0}\bra{0}       & \ket{\sigma_{1}}\bra{\sigma_{1}} &               &               &               \\
                 & \ket{0}\bra{0}   & \ket{\sigma_{2}}\bra{\sigma_{2}} &            &               \\
                 &                  & \ddots        & \ddots        &               \\
                 &                  &               & \ket{0}\bra{0}  & \ket{\sigma_{N}}\bra{\sigma_{N}}\\
                 &                  &               &               & \ket{0}\bra{0}
    \end{pmatrix}_{L}
    \bmm{e}_{N+1}\nt
    &=\bmm{e}_{1}^{\top}
    \begin{pmatrix}
1-\hat{n}_{1}       & \hat{n}_{1,\sigma_{1}} &               &               &               \\
                 & 1-\hat{n}_{1}   & \hat{n}_{1,\sigma_{2}} &            &               \\
                 &                  & \ddots        & \ddots        &               \\
                 &                  &               & 1-\hat{n}_{1}  & \hat{n}_{1,\sigma_{N}} \\
                 &                  &               &               & 1-\hat{n}_{1}
    \end{pmatrix}
    \cdots
    \begin{pmatrix}
1-\hat{n}_{L}       & \hat{n}_{L,\sigma_{1}} &               &               &               \\
                 & 1-\hat{n}_{L}   & \hat{n}_{L,\sigma_{2}} &            &               \\
                 &                  & \ddots        & \ddots        &               \\
                 &                  &               & 1-\hat{n}_{L}  & \hat{n}_{L,\sigma_{N}} \\
                 &                  &               &               & 1-\hat{n}_{L}
    \end{pmatrix}
    \bmm{e}_{N+1}.
\end{align}
\end{widetext}
Applying projection $\hat P\qty(S)$ to $\ket{\qty(r_{1},\dots,r_{M})}$ yields simultaneous eigenstates of $H$ and $\mathcal{M}$.

Counting shows that for fixed $N$, there are
\begin{align}
  \sum_{r_{1}+\cdots+r_{M}=N} \frac{N!}{r_{1}!\cdots r_{M}!} = M^{N}
\end{align}
ground states, and summing $N=0,\dots,L$ gives $\sum_{N=0}^{L}M^{N}=\frac{M^{L+1}-1}{M-1}$, matching Sec.~\ref{subsec:uniqueness}.  

Finally, by contracting zero singular‑value bonds in the MPO form \eqref{eq:BgeneratorMPO} after projection, one recovers the MPS representation \eqref{eq:Psidash}.

\section{Numerical Evidence for the Gap Existence}\label{app:gappedproof}

We apply Knabe’s method~\cite{lemm2019spectral} (originally formulated for open chains) to the present reflecting-boundary no-passing ASEP. 
Each two-site term $h_{j,j+1}$ in \eqref{eq:HermHam} is a projector with eigenvalues $0$ and $1$. Within this framework, if the finite-size gap exceeds the threshold function $G\qty(L)$ in ~\eqref{eq:threshold}, then the model is rigorously gapped in the thermodynamic limit.

\begin{figure}[H]
  \includegraphics[width=\columnwidth,clip]{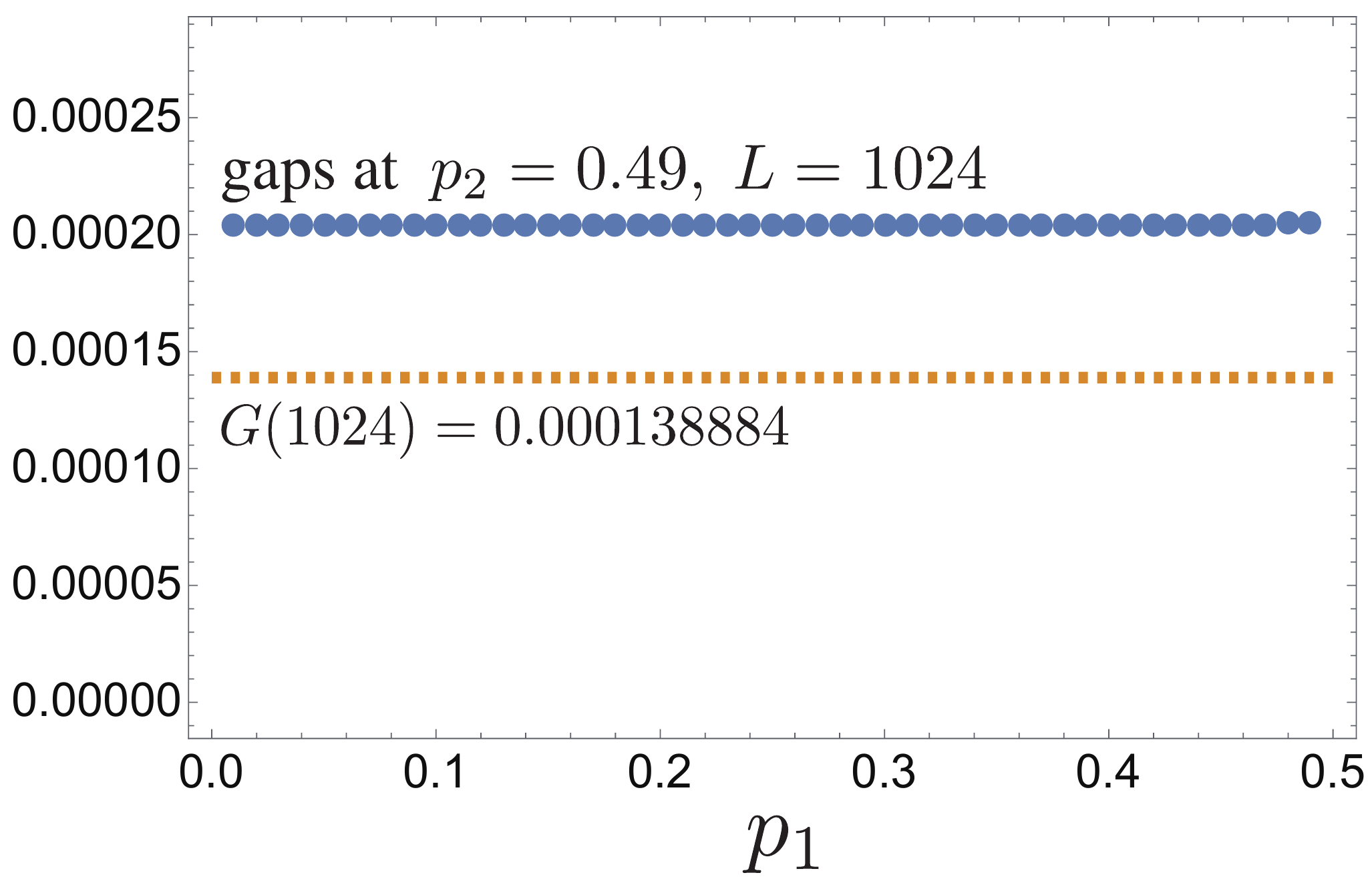}
  \caption{Numerical verification of the gap using Knabe’s method [open-chain formulation, applied here to reflecting (closed) boundary conditions]~\cite{lemm2019spectral}.  
  Blue points show the minimal gap $\Delta_{\mathrm{min}}$ at $L=1024$ for $p_{2}=0.49$ and $p_{1}=0.01,0.02,\cdots,0.49$.  
  The orange dashed line marks the threshold $G\qty(1024)=0.000138884$.  
  All gaps exceed the threshold, confirming these points are gapped.  
  \label{Fig-GapKnabe}}
\end{figure}

In Sec.~\ref{subsec:minimum-gap}, we argued that the red regions in Fig.~\ref{Fig-mingapsector} ($p_{1}<\hf$, $p_{2}<\hf$ or $p_{1}>\hf$, $p_{2}>\hf$) are consistent with gapped behavior.  By the  
parameter-space symmetry of Sec.~\ref{subsec:mirrorUnitary}, it is sufficient to check the lower half of the yellow simplex in Fig.~\ref{Fig-mirror}.  We computed $\Delta_{\mathrm{min}}$ by exact diagonalization at $L=1024$ for boundary points $p_{2}=0.49$ and $p_{1}=0.01,0.02,\cdots,0.49$.  Fig.~\ref{Fig-GapKnabe} compares these values with $G\qty(1024)=0.000138884$.  All computed gaps lie above the threshold, confirming that these sampled boundary points are gapped.

We also checked interior points of the red region by varying $\qty(p_{1},p_{2})$ from $\qty(0.01,0.01)$ in steps of $0.01$ at $L=1024$.  In every sampled case, $\Delta_{\mathrm{min}}>G\qty(1024)$, further supporting an $O(1)$ gap in this region.

\section{Gap Closing Data for the Smallest Gap Sector}\label{app:expsmallnumerical}

In Sec.~\ref{subsec:minimum-gap}, we claimed that for $p_{1}=\hf$ or $p_{2}=\hf$ (the cyan lines in Fig.~\ref{Fig-mingapsector}), the gap $\Delta_{\mathrm{min}}$ closes algebraically with system size $L$.  In contrast, for $p_{1}<\hf<p_{2}$ or $p_{2}<\hf<p_{1}$ (the purple regions in Fig.~\ref{Fig-mingapsector}), $\Delta_{\mathrm{min}}$ closes exponentially with system size $L$.  Here we present numerical evidence supporting these behaviors.

\begin{figure}[H]
  \includegraphics[width=\columnwidth,clip]{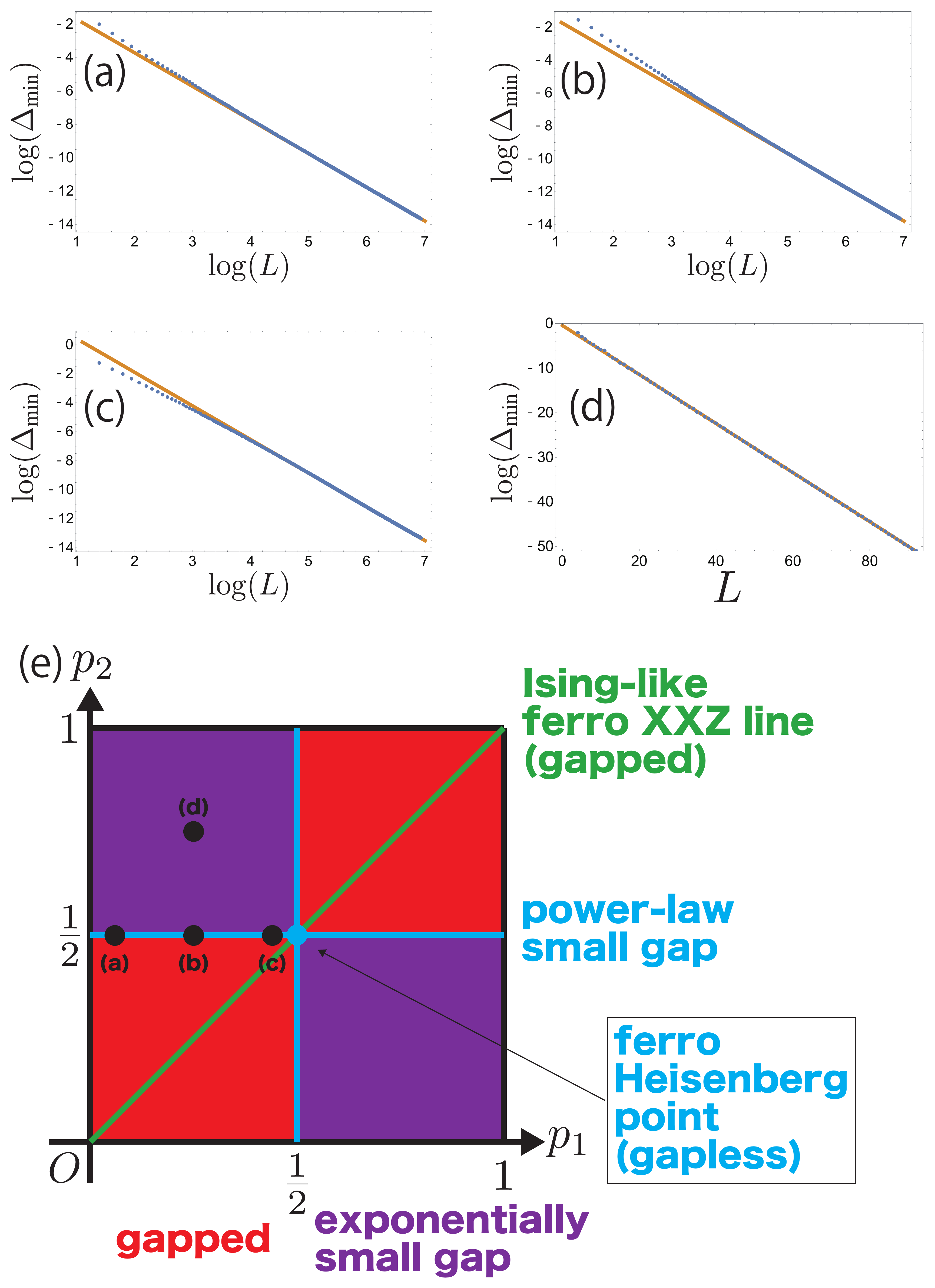}
  \caption{Minimal gap $\Delta_{\mathrm{min}}$ versus system size $L$.  
  (a)–(c): Log–log plots at $\qty(p_{1},p_{2})=\qty(0.01,0.5)$, $\qty(0.25,0.5)$, and $\qty(0.49,0.5)$, showing algebraic gap closing with dynamical critical exponents $z=2.015$, $2.044$, and $2.320$.  
  (d): Semi–log plot at $\qty(p_{1},p_{2})=\qty(0.25,0.75)$, showing exponential gap closing.  
  (e): Positions of these points in the parameter space from Fig.~\ref{Fig-mingapsector}.  
  \label{Fig-expgap}}
\end{figure}

Fig.~\ref{Fig-expgap}\ (a)--(c) show log–log plots of $\Delta_{\mathrm{min}}$ versus $L$ for 
$\qty(p_{1},p_{2}) = \qty(0.01,0.5)$, $\qty(0.25,0.5)$, and $\qty(0.49,0.5)$, respectively.
In all three cases, the data follow a power law and fits give dynamical critical exponents $z=2.015$, $2.044$, and $2.320$.
These $z\ge 2$ results are consistent with the result of Refs.~\cite{PhysRevB.110.195140,masaoka2024rigorouslowerbounddynamic}.
We also checked additional points with $p_{2}=\hf$ and $p_{1}=0.01,0.02,\cdots,0.49$, and observed the same algebraic scaling.

Fig.~\ref{Fig-expgap} (d) is a semi–log plot of $\Delta_{\mathrm{min}}$ versus $L$ at $\qty(p_{1},p_{2})=\qty(0.25,0.75)$.  The linear trend confirms exponential gap closing. Using the symmetry of Sec.~\ref{subsec:mirrorUnitary} and the same $0.01$ grid sampling as in Appendix~\ref{app:gappedproof}, we observed the same exponential trend at all sampled points in the purple region.

\clearpage
\section{Matrix Representation of the Hamiltonian in the Smallest Gap Sector}\label{app:Hamrep}
We give here the matrix form of the Hamiltonian in the smallest gap sector used for numerical calculations in SubSec.~\ref{subsec:minimum-gap} and Appendices~\ref{app:gappedproof} and \ref{app:expsmallnumerical}.  
In this sector, the basis is labeled solely by the position \(j\) of the hole state \(\ket{0}\).  
On that basis, the Hamiltonian matrix is
\begin{align}
\begin{pmatrix}
p_1      & -t_1 &        &        &        &        &        &        &         \\
-t_1   & 1      & \ddots &        &        &        &        &        &         \\
         & \ddots & \ddots & \ddots &        &        &        &        &         \\
         &        & \ddots & 1      & -t_1 &        &        &        &         \\
         &        &        & -t_1 & 1 - p_{1} + p_{2} & -t_2 &        &        &         \\
         &        &        &        & -t_2 & 1      & \ddots &        &         \\
         &        &        &        &        & \ddots & \ddots & \ddots &         \\
         &        &        &        &        &        & \ddots & 1      & -t_2   \\
         &        &        &        &        &        &        & -t_2 & 1 - p_2  
\end{pmatrix},
\end{align}
where 
\begin{align}
    t_{1}&\coloneqq\sqrt{p_{1}\qty(1-p_{1})},\\
    t_{2}&\coloneqq\sqrt{p_{2}\qty(1-p_{2})}.
\end{align}

\end{document}